\documentclass[aps,prb,
superscriptaddress,
 amsmath,amssymb,
preprint,%
]{revtex4-2} 

\usepackage{amsmath}
\usepackage{bm}
\usepackage{amsthm}
\usepackage{amssymb}
\usepackage{mathrsfs}
\usepackage{xcolor}
\usepackage{chemformula}
\usepackage{multirow}
\usepackage[hidelinks]{hyperref}
\usepackage{float}
\usepackage{graphicx} 
\usepackage{tabularx}
\usepackage{physics}
\usepackage{braket}
\usepackage[ruled, linesnumbered]{algorithm2e}
\usepackage[version=4]{mhchem}
\usepackage[caption=false]{subfig}
\usepackage{booktabs}

\begin{document}

\begin{center}
    {\large \textbf{Benchmarking the Impact of Active Space Selection on the VQE Pipeline for Quantum Drug Discovery}}

    \vspace{10pt}

    Zhi Yin$^{1, 3^*}$, 
    Xiaoran Li$^{1,2^*}$,
    Zhupeng Han$^{1}$,
    Shengyu Zhang$^{4}$,
    Xin Li$^{5}$,
    Zhihong Zhang$^{6}$,
    Runqing Zhang$^{6}$,
    Anbang Wang$^{6}$,
    Xiaojin Zhang$^{2^*}$

    
    \textit{$^1$ QureGenAI(aka AceMapAI) Suzhou Biotechnology Co.,Ltd, Suzhou, 215000, China}
    
    \textit{$^2$ QureGenAI Joint Lab, China Pharmaceutical University, Nanjing, 211198, China}
    
    \textit{$^3$ School of Statistics and Data Science, Ningbo University of Technology, Ningbo, 315211, China}
    
    \textit{$^4$ Tencent Quantum Lab, Hong Kong SAR, China}

    \textit{$^5$ Tencent Quantum Lab, Shenzhen, 518057, China}

    \textit{$^6$ China Mobile (Suzhou) Software Technology Company Limited, Suzhou 215163, China}

    \vspace{10pt}

    {\small
    
    $^*$ Corresponding authors. E-mail: yz@nbut.edu.cn; lixiaoran@quregenai.com; zxj@cpu.edu.cn
    }
\end{center}


\begin{abstract}

Quantum computers promise scalable treatments of electronic structure, yet applying variational quantum eigensolvers (VQE) on realistic drug‑like molecules remains constrained by the performance limitations of near‑term quantum hardwares. A key strategy for addressing this challenge which effectively leverages current Noisy Intermediate-Scale Quantum (NISQ) hardwares yet remains under‑benchmarked is active space selection. We introduce a benchmark that heuristically proposes criteria based on chemically grounded metrics to classify the suitability of a molecule for using quantum computing and then quantifies the impact of active space choices across the VQE pipeline for \emph{quantum drug discovery}. The suite covers several representative drug‑like molecules (e.g., lovastatin, oseltamivir, morphine) and uses chemically motivated active spaces. Our VQE evaluations employ both simulation and quantum processing unit (QPU) execution using unitary coupled-cluster with singles and doubles (UCCSD) and hardware-efficient ansatz (HEA). We adopt a more comprehensive evaluation, including chemistry metrics and architecture‑centric metrics. For accuracy, we compare them with classical quantum chemistry methods. This work establishes the first systematic benchmark for active space driven VQE and lays the groundwork for future hardware-algorithm co‑design studies in \emph{quantum drug discovery}.

\end{abstract}

\maketitle

\section{Introduction}

Quantum computing holds transformative promise for computational chemistry and drug discovery, offering the potential to simulate molecular systems with higher accuracy and significantly reduced runtime compared to classical methods~\cite{reiher2017elucidating, bauer2020quantum, grimsley2019adaptive}. Variational quantum eigensolvers (VQE)~\cite{peruzzo2014variational} have emerged as one of the most promising algorithms for near-term quantum hardwares, using a parameterized quantum circuit and a classical optimizer to minimize the energy. Unlike fault-tolerant algorithms such as quantum phase estimation that require millions of logical qubits, VQE can operate on current noisy intermediate-scale quantum (NISQ) hardware with tens to hundreds of physical qubits~\cite{preskill2018quantum}. This practical advantage has motivated extensive efforts to apply VQE to drug-like molecules, where accurate electronic structure calculations are critical for understanding molecular properties, reaction mechanisms, and protein-ligand binding affinities~\cite{santagati2024drug,kandala2017hardware}.

However, applying VQE to realistic drug molecules remains severely constrained 
by the limitations of current quantum hardware. State-of-the-art quantum 
processors feature only $\sim$100--1000 physical qubits with typical single‑ and two‑qubit gate fidelities in the 99.9\% and 99\%-99.9\% ranges, respectively~\cite{kim2023evidence,google2025quantum}, while even moderately sized drug molecules such as lovastatin (C$_{24}$H$_{36}$O$_5$, 65 atoms) or imatinib (C$_{29}$H$_{31}$N$_7$O, 68 atoms) would naively require hundreds of logical qubits to represent their full electronic structure. This exponential scaling of quantum resources with molecular size creates a critical gap between the promise of quantum advantage and the reality of NISQ device capabilities. 

A key strategy to bridge this gap is \emph{active space selection}~\cite{khedkar2019active,de2023complete}, which restricts quantum computation to a chemically relevant subset of molecular orbitals and electrons. By carefully selecting an active space—typically denoted as (n$_e$, n$_o$) for n$_e$ electrons in n$_o$ orbitals—one can dramatically reduce the number of qubits required while retaining essential multi-reference correlation effects. For instance, a (6e, 6o) active space requires only 12 qubits under the Jordan-Wigner mapping~\cite{jordan1928paulische,nielsen2005fermionic}, making VQE feasible on current hardware, whereas the full orbital space of even a small drug molecule would require 50+ qubits. Despite its critical importance, active space selection for electronic structure calculation using quantum computing remains largely heuristic and insufficiently benchmarked. Classical quantum chemistry methods such as complete active space self-consistent field (CASSCF)~\cite{roos1980complete,alessandro2025complete} have established heuristics based on chemical intuition—focusing on frontier orbitals, bonds being broken or formed, and strongly correlated electron pairs. However, these guidelines were developed for classical algorithms and may not directly translate to the VQE pipeline especially on near-term QPU, where circuit depth, measurement overhead, and ansatz expressibility introduce additional constraints. Furthermore, not all molecules benefit equally from quantum computation: single-reference systems can often be handled accurately by density functional theory (DFT)~\cite{hohenberg1964inhomogeneous,jones2015density} or coupled cluster methods~\cite{koch1990coupled,stanton1993equation}, while highly multi-reference systems may require active spaces too large for NISQ devices. Identifying which molecules are suitable for quantum computing, and how to optimally select their active spaces, remains an open question with significant practical implications.

Existing work on VQE for molecular systems has focused mainly on small 
reference molecules such as H$_2$, H$_2$O, LiH and BeH$_2$, or specific 
case studies of individual drug-like molecules~\cite{google2020hartree}. While these studies have demonstrated 
the feasibility of VQE and explored various ansatze (e.g., unitary coupled 
cluster with singles and doubles [UCCSD]~\cite{taube2006new,bartlett2007coupled,anand2022quantum}, 
hardware-efficient ansatze [HEA]~\cite{kandala2017hardware,sim2019expressibility,ostaszewski2021structure,schuld2019evaluating,bharti2022noisy}), they lack 
systematic evaluation across a diverse set of drug-relevant molecules with 
varying sizes and electronic structures. Critically, there exists no 
comprehensive benchmark that (1) establishes heuristic criteria to classify 
molecules' suitability for quantum computing, (2) systematically quantifies 
the impact of active space choices across the VQE pipeline, and (3) evaluates 
performance using both chemistry-centric metrics (such as energy accuracy) and architecture-centric metrics (such as qubit count, circuit depth, gate 
complexity). Such a benchmark is essential for guiding practical applications 
of quantum drug discovery and informing future hardware-algorithm co-design.

In this work, we present the first systematic benchmark for active space 
selection in the VQE pipeline for quantum drug discovery. We make the 
following key contributions:

\noindent\textbf{Heuristic classification criteria.} We introduce chemically 
grounded metrics to assess a molecule's suitability for quantum computing 
based on natural orbital occupation numbers derived from CASSCF calculations. 
By analyzing fractional occupations in a standardized active space probe, 
we rapidly classify molecules by multi-reference character strength, enabling 
pre-screening to identify high-value targets for quantum computation.

\noindent\textbf{Diverse molecular benchmark suite.} We curate a representative set of seven drug-like molecules(with water and benzene as references) spanning 3 to 68 atoms, including clinically important compounds such as aspirin, oseltamivir, morphine, lovastatin, and imatinib. This suite covers a spectrum from single-reference systems (where quantum computing may be unnecessary) to challenging multi-reference cases (where quantum advantage is expected but active spaces may exceed NISQ capabilities), providing a realistic testbed for evaluating active space strategies.

\noindent\textbf{Systematic active space evaluation.} For each molecule, we evaluate multiple active space configurations ranging from minimal (2e, 2o) to extended (4e, 4o) or larger, examining their impact on both accuracy and quantum resource requirements. We employ both UCCSD and HEA ansatze, and compare VQE results against classical benchmarks including Hartree-Fock and density functional theory (DFT).

\noindent\textbf{Multi-dimensional evaluation.} Unlike prior studies that focus solely on energy accuracy, we adopt a comprehensive evaluation encompassing: (i) chemistry metrics: absolute and relative energy errors, comparison with gold-standard methods; (ii) architecture metrics: qubit counts, circuit depth. This multi-faceted analysis reveals trade-offs between chemical fidelity and hardware feasibility.

\noindent\textbf{Actionable insights for hardware-algorithm co-design.} Our benchmark including end-to-end optimization on two distinct superconducting quantum processors (13-qubit homebrew s2 and 60-qubit homebrew q1 devices) identifies which molecular characteristics (size, multi-reference character, electronic structure complexity) correlate with successful VQE performance under real hardware noise. The direct comparison between QPU architectures reveals critical trade-offs: larger processors exhibit increased convergence oscillations (e.g., benzene's fluctuations on q1 vs. smooth descent on s2) yet achieve comparable final energies, demonstrating that there is possibility that current VQE applications are bottlenecked by ansatz depth rather than qubit count. These hardware-validated findings provide concrete guidance for: (i)~computational chemists deciding when to employ quantum computing and which QPU architectures suit specific molecular problems; (ii)~algorithm developers optimizing VQE protocols for noise resilience and convergence stability on real devices; and (iii)~hardware engineers prioritizing improvements—our results suggest gate fidelity and circuit depth limits dominate over raw qubit count for near-term devices, shifting design priorities toward error mitigation and connectivity over scaling alone.

We openly release our benchmark suite, including molecular geometries, active space configurations, classical reference data, and VQE evaluation scripts, to facilitate reproducibility and community-driven extensions.

The remainder of this paper is organized as follows. Section~2 provides 
background on VQE, active space methods, and reviews related work in 
quantum computational chemistry. Section~3 describes our methodology, 
including the classification criteria, active space selection strategies, 
VQE pipeline implementation, and evaluation metrics. Section~4 presents comprehensive results and analysis across our benchmark. Section~5 discusses implications for quantum drug discovery, limitations of the current study, and future directions. Section~6 concludes.



\section{background and related work}
\subsection{Variational Quantum Eigensolver}
\begin{figure*}[h]
    \includegraphics[width=0.9\textwidth]{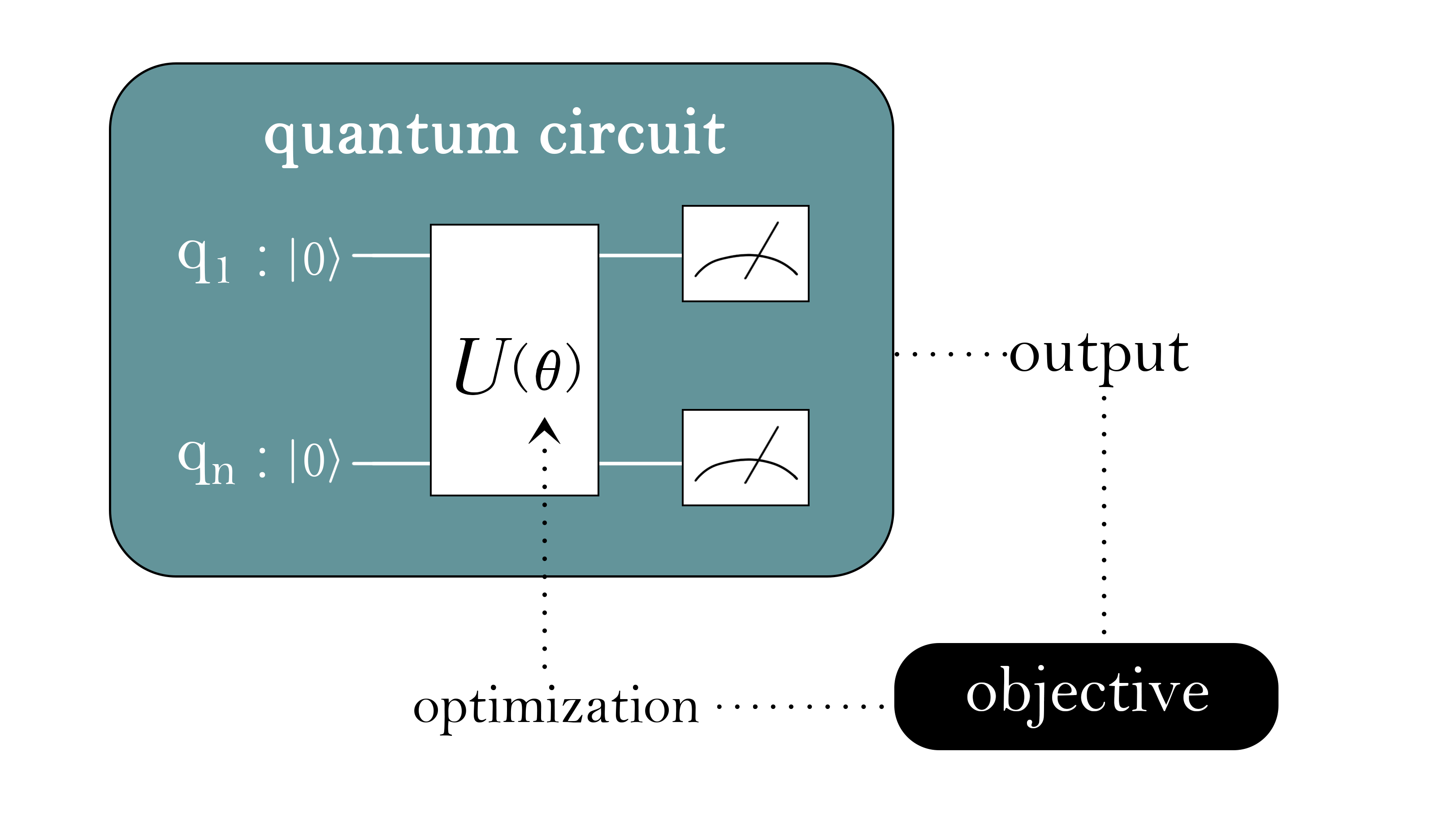}
    \caption{Variational Quantum Eigensolver schematic: parameterized quantum 
circuit $U(\theta)$ generates trial states, measurement outcomes define the 
energy objective, and classical optimization refines $\theta$ iteratively 
until convergence.}
    \label{fig:vqe}
\end{figure*}

The variational quantum eigensolver (VQE) is a hybrid 
quantum-classical algorithm designed to find ground state energies of quantum 
systems(see Figure \ref{fig:vqe}). VQE can be operated on noisy intermediate-scale quantum (NISQ) devices by combining shallow quantum circuits with classical optimization. Given a molecular Hamiltonian $\hat{H}$ represented as a sum of Pauli operators,
\begin{equation}
\hat{H} = \sum_{i} h_i \hat{P}_i,
\end{equation}
where $\hat{P}_i \in \{I, X, Y, Z\}^{\otimes n}$ are $n$-qubit Pauli strings, 
VQE seeks the ground state energy $E_0 = \min_{\psi}\langle\psi|\hat{H}|\psi\rangle$ 
via the variational principle. The algorithm prepares a parameterized quantum 
state $|\psi(\boldsymbol{\theta})\rangle$ using an ansatz circuit $U(\boldsymbol{\theta})$ acting on a reference state (typically Hartree-Fock):
\begin{equation}
|\psi(\boldsymbol{\theta})\rangle = U(\boldsymbol{\theta}) |\text{HF}\rangle.
\end{equation}
The energy expectation value $E(\boldsymbol{\theta}) = \langle\psi(\boldsymbol{\theta})|
\hat{H}|\psi(\boldsymbol{\theta})\rangle$ is evaluated on the quantum processor 
by measuring each Pauli term, and a classical optimizer iteratively updates 
$\boldsymbol{\theta}$ to minimize $E(\boldsymbol{\theta})$.

VQE's shallow circuit depth (compared to quantum phase estimation) makes it tolerant to gate errors and decoherence. The variational bound $E(\boldsymbol{\theta}) \geq E_0$ ensures that even imperfect optimization yields an upper bound on the ground state energy. However, VQE faces challenges including barren plateaus (vanishing gradients in large circuits)~\cite{mcclean2018barren,holmes2022connecting}, sensitivity to local minima, and measurement overhead scaling with the number of Hamiltonian terms.

\subsection{Active Space Approximation}

For a molecule with $N$ electrons in $M$ spatial orbitals, the full 
configuration interaction (FCI) wavefunction involves $\binom{2M}{N}$ 
determinants, leading to exponential scaling. On a quantum computer using 
Jordan-Wigner, this requires $2M$ qubits—prohibitively 
large for drug molecules.

\begin{figure*}[h]
    \includegraphics[width=0.9\textwidth]{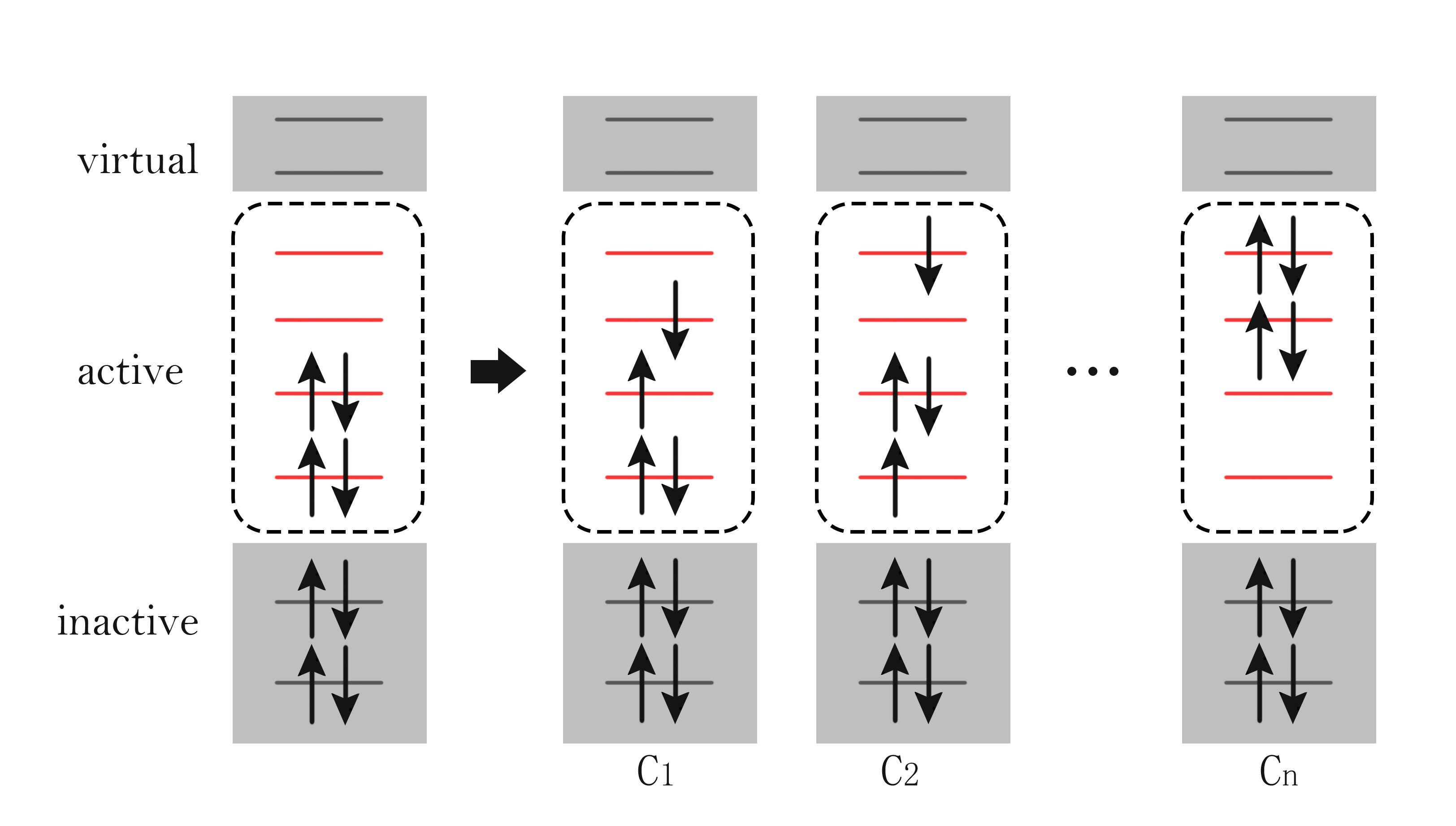}
    \caption{Active space approximation for molecular orbital partitioning. 
Virtual orbitals (top, gray) and inactive orbitals (bottom, gray) remain 
frozen at Hartree-Fock occupation, while electrons in the active space 
(middle, dashed box) explore all possible configurations $C_1, C_2, \ldots, C_n$ 
via quantum computation. Red lines indicate chemically relevant orbitals 
treated variationally; arrows represent electron occupations.}
    \label{fig:active}
\end{figure*}

The active space approximation partitions molecular 
orbitals into three classes(see Figure \ref{fig:active}):
\begin{itemize}
\item \textbf{Core (inactive) orbitals}: Doubly occupied in all configurations, 
      treated at the mean-field level.
\item \textbf{Active orbitals}: Partially occupied orbitals where electron 
      correlation is explicitly treated. An $(n_e, n_o)$ active space contains 
      $n_e$ active electrons distributed among $n_o$ active orbitals.
\item \textbf{Virtual orbitals}: Unoccupied orbitals excluded from the 
      correlation treatment.
\end{itemize}

Typically, active orbitals are chosen around the Fermi level (HOMO-LUMO region), 
focusing on valence electrons involved in bonding, reactions, or multi-reference 
correlation. This reduces the quantum resource requirement to $2n_o$ qubits 
under Jordan-Wigner mapping, making VQE tractable for current hardware when 
$n_o$ is small.

Classical quantum chemistry employs heuristics such as:
\begin{itemize}
\item Selecting orbitals with natural occupation numbers deviating significantly 
      from 2 (occupied) or 0 (virtual).
\item Including orbitals involved in bond breaking/formation.
\item Using localized orbital analysis (e.g., intrinsic bonding orbitals).
\end{itemize}
However, these guidelines were developed for classical multi-reference methods 
(CASSCF~\cite{roos1980complete}, CASPT2~\cite{andersson1992second}, DMRG~\cite{white1992density}) and may not directly translate to VQE, where circuit 
depth, measurement overhead, and ansatz expressibility introduce additional 
constraints. Systematic benchmarking of active space choices for VQE in 
drug-like molecules remains scarce.

\subsection{VQE Ansatze for Quantum Chemistry}

\paragraph{Unitary Coupled Cluster (UCC).}
The coupled cluster ansatz, a gold standard in classical quantum chemistry, 
can be adapted to quantum computers via unitary coupled cluster with singles 
and doubles (UCCSD):
\begin{equation}
|\psi_{\text{UCCSD}}\rangle = e^{\hat{T} - \hat{T}^\dagger} |\text{HF}\rangle,
\end{equation}
where $\hat{T} = \hat{T}_1 + \hat{T}_2$ includes single ($\hat{T}_1$) and 
double ($\hat{T}_2$) excitation operators. The anti-Hermitian form $\hat{T} - \hat{T}^\dagger$ 
ensures unitarity. UCCSD has $O(N_{\text{occ}} N_{\text{virt}} + N_{\text{occ}}^2 N_{\text{virt}}^2)$ 
parameters, providing high chemical accuracy but requiring deep circuits with 
many CNOT gates.

\paragraph{Hardware-efficient ansatze (HEA).}
To reduce circuit depth, hardware-efficient ansatze employ layers of single-qubit 
rotations and entangling gates native to the quantum processor~\cite{kandala2017hardware}:
\begin{equation}
U_{\text{HEA}}(\boldsymbol{\theta}) = \prod_{\ell=1}^{L} \left[ U_{\text{ent}}^{(\ell)} 
\prod_{i} R_y(\theta_i^{(\ell)}) R_z(\phi_i^{(\ell)}) \right],
\end{equation}
where $U_{\text{ent}}^{(\ell)}$ is the entangling layer for the $\ell$-th circuit 
layer (e.g., linear CX gates). HEA 
circuits are shallow and noise-resilient but lack direct chemical interpretability. The trade-off between UCCSD (high accuracy, deep circuits) and HEA (lower accuracy, shallow circuits) is crucial for NISQ applications.

\subsection{Quantum Drug Discovery}

Early VQE demonstrations focused on small benchmark molecules: H$_2$, H$_2$O, LiH, and BeH$_2$~\cite{peruzzo-2014,omalley-2016,kandala2017hardware}. 
Recent efforts have targeted larger systems, including simulations of drug-like 
molecules. For instance, research~\cite{li2024hybrid}reported VQE calculations for real-world 
drug design challenges, specifically computing Gibbs free energy profiles 
for prodrug activation and simulating covalent bond interactions.

However, most studies are \emph{case-specific} rather than systematic benchmarks. 
They typically:
\begin{itemize}
\item Focus on a single molecule or narrow chemical class.
\item Report results for one or two active space configurations without 
      exploring sensitivity or convergence.
\item Evaluate only simulator-based VQE, lacking real QPU validation.
\end{itemize}

Furthermore, no prior work has established \emph{heuristic criteria} to classify molecules by their suitability for quantum computing, nor systematically quantified the impact of active space choices across a diverse set of drug-relevant molecules with varying sizes and electronic structures. Our benchmark addresses these gaps by providing:
\begin{enumerate}
\item A classification scheme based on multi-reference character.
\item Systematic active space evaluation for 7 representative drug molecules.
\item Multi-dimensional assessment encompassing both chemistry metrics (energy       accuracy) and architecture metrics (qubit count, circuit depth).
\item Validation on physical quantum processors.
\end{enumerate}

This establishes the first comprehensive benchmark for active space-driven VQE in quantum drug discovery, providing actionable guidance for practitioners.

\section{methodology}


\begin{figure*}[h]
    \includegraphics[width=1\textwidth]{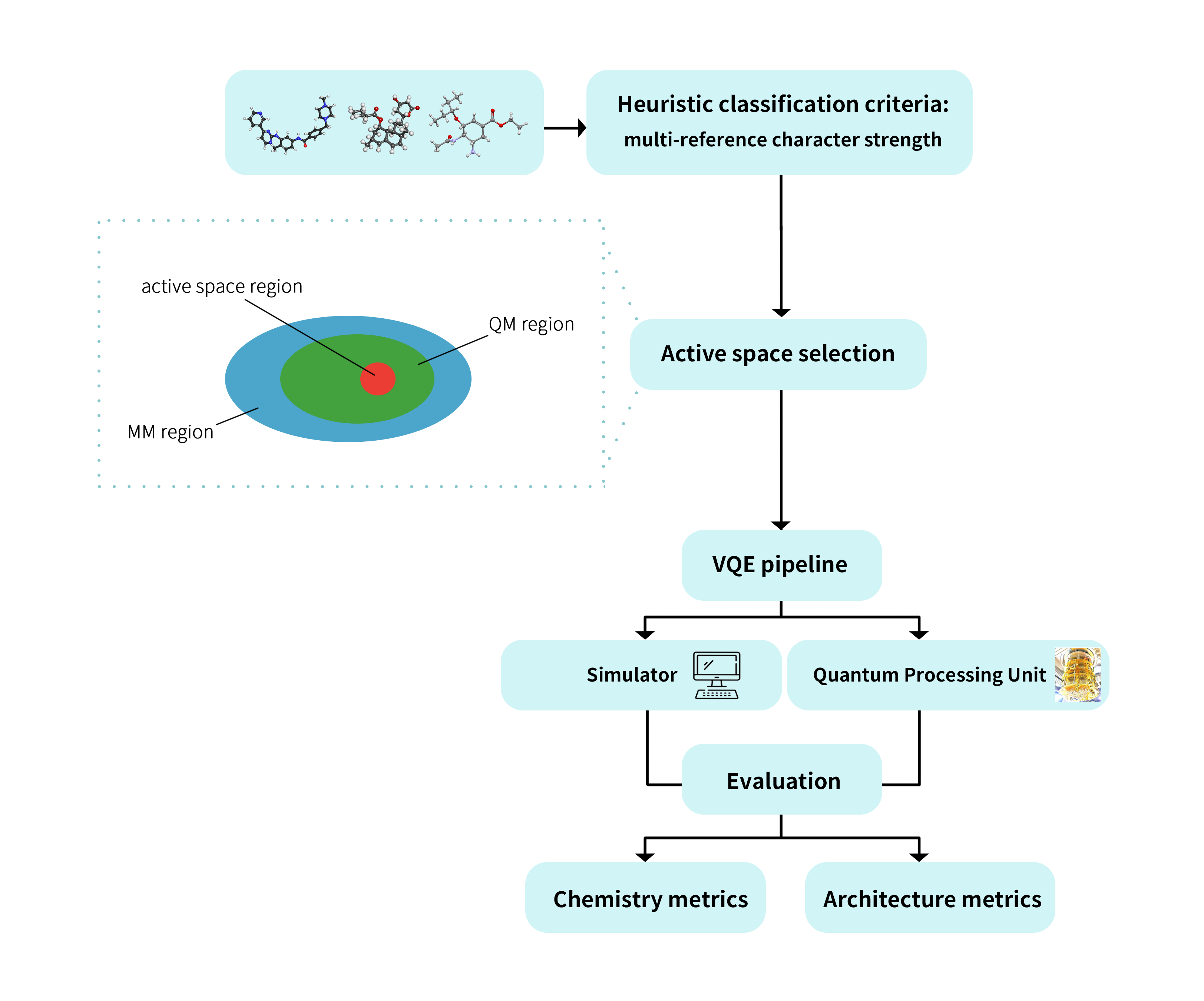}
    \caption{Our VQE benchmark workflow. Molecules undergo heuristic classification for quantum computing suitability, followed by active space selection(left: multiscale framework for molecules calculation, QM=quantum mechanics, MM=molecualr mechanics). VQE calculations execute on simulators and QPUs, with evaluation via chemistry metrics (energy accuracy) and architecture metrics (circuit resources). This systematic pipeline enables cross-platform assessment of VQE performance for drug-like molecules.}
    \label{fig:workflow}
\end{figure*}


Our methodology comprises four stages: (i)~heuristic classification to assess 
molecular suitability for quantum computing; (ii)~active space selection strategies tailored for VQE; (iii)~VQE pipeline implementation on both simulators and quantum processors; and (iv)~multi-dimensional evaluation metrics. (Figure \ref{fig:workflow} illustrates the overall workflow)

\subsection{Molecule Classification via Multi-reference Diagnostics}

To rapidly identify which drug-like molecules are promising candidates for quantum 
computing, we establish a heuristic classification scheme grounded in electronic 
structure theory.

Not all molecules require quantum computational treatment. Single-reference systems, where electron correlation is weak and mean-field (Hartree-Fock) or perturbative (DFT, MP2) methods suffice, gain little from VQE's multi-reference capabilities. Conversely, strongly correlated systems—where near-degenerate states and fractional orbital occupations dominate—challenge classical methods but are natural targets for quantum algorithms. A pre-screening classification enables efficient resource 
allocation.

\subsubsection{Multi-reference Character Assessment}

We employ complete active space self-consistent field (CASSCF) calculations with 
a uniform $(6e, 6o)$ active space as a standardized probe across all molecules. 
This choice is motivated by:
\begin{itemize}
\item \textbf{Hardware accessibility}: A $(6e, 6o)$ active space maps to 12 qubits 
      under Jordan-Wigner encoding, placing it within reach of current NISQ devices.
\item \textbf{Frontier orbital coverage}: Six orbitals typically encompass 
      HOMO$-$2 through LUMO$+$2, capturing valence correlation relevant to chemical 
      bonding and reactivity.
\item \textbf{Standardized comparison}: Using a fixed active space across molecules 
      of varying sizes enables fair assessment of multi-reference character, analogous 
      to employing a consistent basis set in benchmark studies.
\end{itemize}

We acknowledge that $(6e, 6o)$ does not represent the optimal active space for 
each individual molecule. Smaller systems like H$_2$O are over-represented (60\% 
of electrons), while larger systems like imatinib may under-capture correlation 
effects (only 9\% of electrons). However, for \emph{initial classification and 
ranking}, this uniform approach provides a computationally tractable and 
interpretable metric.

\subsubsection{Natural Orbital Occupation Numbers}

Given a CASSCF$(6,6)$ calculation, we analyze the natural orbital occupation 
numbers $\{n_i\}$ obtained by diagonalizing the one-particle reduced density 
matrix over the active space:
\begin{equation}
\hat{\gamma} = \sum_{i=1}^{6} n_i |\phi_i\rangle\langle\phi_i|, \quad 
\sum_{i} n_i = 6 \text{ (electrons)}.
\end{equation}
In a pure single-reference system, $n_i \approx 2$ for occupied orbitals and 
$n_i \approx 0$ for virtual orbitals. Strong multi-reference character manifests 
as fractional occupations deviating significantly from these limits.

We define a natural orbital as \textbf{fractionally occupied} if:
\begin{equation}
0.1 < n_i < 1.9
\end{equation}
This threshold range, informed by prior CASSCF studies~\cite{sayfutyarova-2017}, reflects meaningful departure from integer occupation. 
The number of fractionally occupied orbitals, $N_{\text{frac}}$, quantifies 
multi-reference strength.

\subsubsection{Classification Scheme}

Based on $N_{\text{frac}}$, we classify molecules into three categories:

\begin{itemize}
\item \textbf{Strong multi-reference} ($N_{\text{frac}} \geq 4$): At least four orbitals exhibit fractional occupation, indicating substantial static correlation. These systems typically challenge single-reference methods (DFT, CCSD) and demand multi-reference treatments (CASPT2, DMRG) that scale poorly. \emph{Prime candidates for quantum computing.}

\item \textbf{Moderate multi-reference} ($2 \leq N_{\text{frac}} < 4$): Two to three orbitals show fractional character, suggesting non-negligible but moderate correlation effects. Quantum computing may offer advantages, particularly for subtle correlation contributions difficult for DFT. \emph{Good candidates, benefit likely observable.}

\item \textbf{Weak/Single-reference} ($N_{\text{frac}} < 2$): Fewer than two fractionally occupied orbitals, indicating correlation effects are well-localized. Classical methods (DFT, coupled cluster) are typically sufficient. \emph{Quantum computing may be unnecessary overhead.}
\end{itemize}

We emphasize that these thresholds are heuristic guidelines rather than rigorous 
boundaries. The primary goal is to stratify our molecular suite into distinct 
regimes for prioritizing subsequent VQE evaluations.

\subsubsection{Limitations and Future Refinements}

Our classification has several acknowledged limitations:
\begin{enumerate}
\item \emph{Active space uniformity}: The fixed $(6,6)$ choice may not optimally represent all molecules. Future work should explore molecule-adaptive selection guided by localization analysis or iterative schemes.
      
\item \emph{Occupation threshold sensitivity}: The $0.1$--$1.9$ range and thresholds $(2, 4)$ are heuristic. More sophisticated metrics could weight occupations by deviation magnitude or incorporate entropy-based measures.
      
\item \emph{Basis set effects}: We employ minimal STO-3G basis for computational expediency. While suitable for classification, quantitative occupation numbers and energies may shift with larger basis sets.
\end{enumerate}

Despite these limitations, our classification provides a rapid, interpretable, 
and hardware-cognizant screening tool validated by subsequent VQE benchmarks.

\subsection{Active Space Selection Strategies}
\label{sec:as-selection}

Having classified molecules, we design active space strategies for VQE evaluation 
that balance chemical accuracy with quantum resource constraints.

\subsubsection{Guiding Principles}

Active space selection for VQE must consider factors beyond classical criteria:

\begin{itemize}
\item \textbf{Qubit count}: $(n_e, n_o)$ maps to $2n_o$ qubits (Jordan-Wigner),       directly determining hardware feasibility.      

\item \textbf{Circuit depth}: Larger active spaces increase UCCSD parameter count $\propto n_e(2n_o - n_e)$, deepening circuits and amplifying gate errors.

\item \textbf{Measurement overhead}: Hamiltonian terms scale as $O(n_o^4)$,       inflating measurement shots required for accurate energy estimation.
\end{itemize}

Our strategy systematically varies active space size to map the accuracy-resource trade-off curve for each molecule.

\subsubsection{Active Space Configurations}

For each molecule, we evaluated multiple active space sizes:

\begin{itemize}
\item \textbf{Minimal}: $(2e, 2o)$ — 4 qubits. Captures HOMO and LUMO, the most essential frontier orbitals. Applicable to smallest molecules (H$_2$O) and serves as proof-of-concept for QPU validation. Represents the lower bound on correlation treatment.

\item \textbf{Standard}: $(4e, 4o)$ — 8 qubits. Encompasses HOMO$-$1, HOMO, LUMO, LUMO$+$1, capturing primary valence correlation. This active space size represents a practical target for current NISQ devices, balancing chemical accuracy with hardware feasibility (8 qubits are well within capabilities of state-of-the-art quantum processors).
\end{itemize}

We focus on these NISQ-accessible active space sizes rather than larger 
configurations (e.g., $(6e, 6o)$, $(8e, 8o)$) for alignment with realistic near-term quantum hardware capabilities. Larger active spaces, while chemically desirable for capturing additional dynamic correlation, would require 12--16 qubits and substantially deeper circuits, exceeding the practical limits of current noisy devices. Our benchmark thus intentionally targets the \emph{accessible frontier} of 
NISQ-era quantum drug discovery.

Active orbitals are selected from canonical Hartree-Fock orbitals centered around the Fermi level. For an $(n_e, n_o)$ space with $n_{\text{occ}} = n_e/2$ occupied orbitals, we select orbitals:

\begin{equation}\{\text{HOMO}-(n_{\text{occ}}-1), \ldots, \text{HOMO},  \text{LUMO}, \ldots, \text{LUMO}+(n_o - n_{\text{occ}} - 1)\}.
\end{equation}

\subsubsection{Molecular Coverage Strategy}

Given the focus on NISQ-accessible active spaces, we evaluate:
\begin{itemize}

\item \textbf{All molecules}: $(2e, 2o)$ and $(4e, 4o)$ — the primary configuration 
      balancing accuracy and feasibility.

\item \textbf{Ansatz comparison}: UCCSD and HEA for selected drug-like molecules to assess circuit depth vs. accuracy trade-offs.
\end{itemize}

This configuration matrix yields approximately 20 VQE evaluations, 
sufficient to address our benchmark objectives while remaining computationally 
tractable.

\subsection{VQE Pipeline Implementation}

\subsubsection{Simulator-based VQE}

\paragraph{Hamiltonian construction.}
For a given molecule and $(n_e, n_o)$ active space, we construct the electronic 
Hamiltonian using PySCF~\cite{sun-2017} and Qiskit~\cite{qiskit2024}:
\begin{enumerate}
\item Perform restricted Hartree-Fock (RHF) calculation with STO-3G basis to obtain molecular orbitals $\{\phi_i\}$.
\item Apply active space transformation to extract one- and two-electron integrals $(h_{pq}, g_{pqrs})$ over active orbitals.
\item Map fermionic operators to qubit operators via Jordan-Wigner transformation:

\begin{equation}
      \hat{H} = \sum_{pq} h_{pq} \hat{a}_p^\dagger \hat{a}_q + 
      \frac{1}{2}\sum_{pqrs} g_{pqrs} \hat{a}_p^\dagger \hat{a}_q^\dagger 
      \hat{a}_r \hat{a}_s \rightarrow \sum_i h_i \hat{P}_i.
      \end{equation}
\end{enumerate}

The resulting Hamiltonian contains $O(n_o^4)$ Pauli terms, each requiring 
separate measurement.

\paragraph{Ansatz circuits.}
We evaluate two ansatze:

\begin{itemize}
\item \textbf{UCCSD (Unitary Coupled Cluster Singles and Doubles)}: Applies the unitary operator
\begin{equation}
U_{\text{UCCSD}}(\boldsymbol{\theta}) = \exp\left(\sum_{ia} \theta_i^a 
(\hat{a}_a^\dagger \hat{a}_i - \text{h.c.}) + \sum_{ijab} \theta_{ij}^{ab} 
(\hat{a}_a^\dagger \hat{a}_b^\dagger \hat{a}_j \hat{a}_i - \text{h.c.})\right)
\end{equation}
to the Hartree-Fock state $|\text{HF}\rangle$. The parameter count is
\begin{equation}
N_{\text{params}}^{\text{UCCSD}} = n_{\text{occ}}n_{\text{virt}} + 
\frac{1}{4}n_{\text{occ}}^2 n_{\text{virt}}^2,
\end{equation}
UCCSD offers high chemical accuracy but deep circuits (CNOT count $\propto N_{\text{params}}^2$).

\item \textbf{HEA (Hardware-Efficient Ansatz)}: 

Employs $L=1$ layer of single-qubit rotations and linear entanglement:
\begin{equation}
U_{\text{HEA}}(\boldsymbol{\theta}) = 
\left[\prod_{i=1}^{2n_o} R_y(\theta_{i}) R_z(\phi_{i})\right] 
\prod_{i=1}^{2n_o-1} \text{CX}_{i,i+1}.
\end{equation}
Parameter count is $N_{\text{params}}^{\text{HEA}} = 2 \cdot 2n_o = 4n_o$, 
e.g., 16 for $(4,4)$. HEA circuits are shallow (fixed depth regardless of active space) but lack chemical structure, potentially requiring more parameters or additional layers for convergence.
\end{itemize}

\paragraph{Optimization.}
We employ the SLSQP (Sequential Least Squares Programming) optimizer, a gradient-based method suitable for VQE's smooth energy landscapes. The key settings are:
\begin{itemize}
\item Maximum iterations: 100 (sufficient for small-to-medium active spaces).
\item Convergence threshold: $10^{-6}$ Ha in energy change.
\item Initial parameters: zeros for UCCSD (starting from Hartree-Fock limit); 
      small random values for HEA.
\end{itemize}

Energy gradients are approximated via finite differences in our simulator. Each VQE run records final energy, iteration count, and optimization time.

\paragraph{Simulation backend.}
We use TensorCircuit~\cite{zhang-2023}, TenCirChem~\cite{li-2023} and Qiskit~\cite{qiskit2024} for simulation, especially Qiskit's \texttt{StatevectorSimulator} for exact 
wavefunction evolution without shot noise. This provides idealized VQE results, 
establishing upper bounds on achievable accuracy before hardware constraints 
are applied.

\subsubsection{Quantum Processing Unit Validation}

To complement simulator results, we perform targeted validation on two physical 
quantum processors: \textbf{homebrew s2 device}, a 13-qubit superconducting processor and \textbf{homebrew q1 device}, a 60-qubit superconducting processor.
Both QPUs support native execution of single-qubit rotations (Ry, Rz) and 
CNOT gates, suitable for VQE circuit compilation.

\subsection{Evaluation Metrics}

We adopt a multi-dimensional evaluation framework encompassing chemical accuracy, 
quantum resource requirements, and hardware.

\subsubsection{Baseline Methods}

To contextualize VQE performance, we compute reference energies using established 
quantum chemistry methods:

\begin{itemize}
\item \textbf{Hartree-Fock (HF)}: Mean-field approximation, serves as the 
      variational upper bound and VQE starting point. Computed with PySCF/STO-3G.

\item \textbf{Density Functional Theory (DFT)}: B3LYP/6-31G* calculations 
      provide the "industry standard" for drug discovery. DFT energies and 
      geometries (obtained via Gaussian calculations) establish classical 
      baseline accuracy and computational cost.
      
\end{itemize}

\subsubsection{Chemistry Metrics}

\paragraph{Energy accuracy.}
Primary metric: absolute energy error relative to the best available reference,
\begin{equation}
\Delta E = |E_{\text{VQE}} - E_{\text{ref}}|,
\end{equation}
where $E_{\text{ref}}$ is DFT. Chemical 
accuracy threshold is 1.6~mHa ($\approx$1~kcal/mol).

\paragraph{Correlation energy recovery.}
Fraction of correlation energy captured by VQE,
\begin{equation}
f_{\text{corr}} = \frac{E_{\text{VQE}} - E_{\text{HF}}}{E_{\text{ref}} - E_{\text{HF}}}.
\end{equation}
Ideal VQE achieves $f_{\text{corr}} \approx 1$.

\paragraph{Comparison with DFT.}
For practical relevance, we compare VQE against DFT (the method chemists 
actually use):
\begin{equation}
\Delta E_{\text{DFT}} = |E_{\text{VQE}} - E_{\text{DFT}}|.
\end{equation}
If VQE does not significantly outperform DFT (faster and more accessible), 
its practical utility is questionable.

\subsubsection{Architecture Metrics}

\paragraph{Quantum resource requirements.}
\begin{itemize}
\item \textbf{Qubit count}: $2n_o$ (Jordan-Wigner).
\item \textbf{Parameter count}: $N_{\text{params}}$ for UCCSD or HEA.
\item \textbf{Circuit depth}: Total gate count and critical-path depth (affects 
      error accumulation).
\item \textbf{CNOT count}: Two-qubit gates dominate error rates on current hardware.
\end{itemize}

\paragraph{Computational cost.}
\begin{itemize}
\item \textbf{Simulator time}: Wall-clock time for VQE convergence (includes 
      energy evaluations and optimization overhead).
\item \textbf{QPU time}: Queue time + execution time on physical devices.
\item \textbf{DFT time}: Comparison baseline for practical feasibility.
\end{itemize}
\begin{table}[h]
\centering
\caption{Experimental configuration matrix. All molecules evaluated at $(2e, 2o)$ and $(4e, 4o)$ active space (4 and 8 qubits, respectively), balancing chemical accuracy with NISQ device accessibility.}
\label{tab:exp-matrix}
\small
\begin{tabular}{llllll}
\toprule
Molecule  & Active Space & Ansatze & Backend & Priority \\
\midrule
H$_2$O          & (2,2)       & UCCSD, HEA       & Sim+QPU† & Medium \\
              &(4,4)           &        &        &       & \\
Aspirin           & (2,2)       & UCCSD, HEA  & Sim+QPU† & High \\
    &(4,4)           &        &        &       & \\
Benzene        & (2,2)       & UCCSD, HEA  & Sim+QPU† & High \\
    &(4,4)           &        &        &       & \\
Oseltamivir    & (2,2)       & UCCSD, HEA  & Simulator & Highest \\
    &(4,4)           &        &        &       & \\
Morphine       & (2,2)       & UCCSD, HEA  & Simulator & Highest \\
    &(4,4)           &        &       &      & \\
Lovastatin     & (2,2)       & UCCSD, HEA  & Simulator      & Medium \\
    &(4,4)           &        &      &       & \\
Imatinib       & (2,2)       & UCCSD, HEA  & Simulator      & Low \\
    &(4,4)           &        &        &       & \\
\bottomrule
\multicolumn{6}{l}{\small † Full VQE optimization on QPU} \\

\end{tabular}
\end{table}

\subsection{Experimental Configurations}

Table~\ref{tab:exp-matrix} summarizes the full experimental matrix, detailing 
which molecules, active spaces, ansatze, and backends are evaluated. The 
complete configuration spans approximately 25--30 VQE tasks for simulator 
and $\sim$10 tasks for QPU validation.

All calculations in simulator baseline basis employ STO-3G for computational tractability. While this minimal basis limits quantitative accuracy, it is adequate for comparative benchmarking and resource scaling analysis. To assess basis set dependence, we performed QPU evaluations using both the minimal STO-3G basis and the polarized 6-31G(d) basis.


\section{Results and Analysis}

\subsection{Molecular Classification and Multi-reference Characterization}
\begin{figure*}[h]
    \includegraphics[width=1\textwidth]{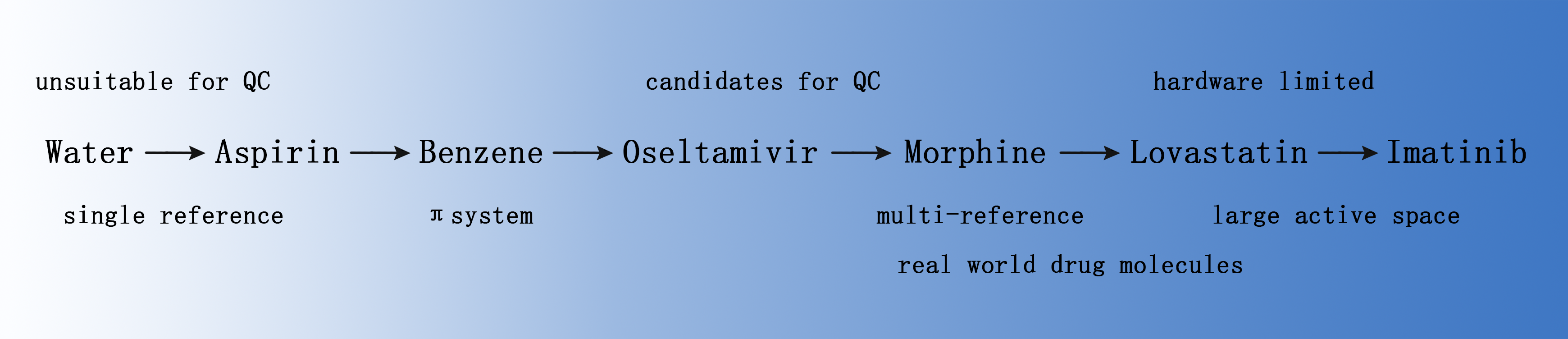}
    \caption{Molecular suitability spectrum for quantum computing. Seven molecules 
arranged from unsuitable (left) to hardware-limited (right) based on quantum 
computing applicability. Upper gradient: H$_2$O (single-reference benchmark) 
$\rightarrow$ oseltamivir/morphine (optimal candidates with manageable 
multi-reference character) $\rightarrow$ lovastatin/imatinib 
(hardware-limited by large active space requirements). Lower annotations 
indicate molecular properties—single-reference nature, $\pi$-system complexity, 
multi-reference character, active space size—that determine VQE suitability. 
This gradient captures the practical window where quantum advantage emerges 
for drug-like molecules.}
    \label{fig:spec}
\end{figure*}

We first classify the seven drug-like molecules(see Figure \ref{fig:spec}) using CASSCF(6,6) natural orbital occupation analysis as described in Section~\ref{sec:as-selection}. Table~\ref{tab:classification} summarizes the multi-reference character assessment for our molecular suite.

\begin{table}[h]
\centering
\caption{Molecular classification based on CASSCF(6,6) natural orbital 
occupation numbers. N$_{\text{frac}}$ denotes the number of fractionally 
occupied orbitals.}
\label{tab:classification}
\small
\begin{tabular}{lccccc}
\toprule
Molecule & N$_e$ & N$_{atoms}$ & N$_{\text{frac}}$ & Score & Category \\
\midrule
H$_2$O         & 10  & 3  & 0 & 4 & Weak \\
Benzene        & 42  & 12 & 0 & 4 & Weak \\
Aspirin        & 94  & 21 & 0 & 4 & Weak \\
Morphine       & 152 & 40 & 0 & 4 & Weak \\
Lovastatin     & 220 & 65 & 0 & 4 & Weak \\
\midrule
Oseltamivir    & 170 & 50 & 2 & 7 & Moderate \\
Imatinib       & 262 & 68 & 2 & 7 & Moderate \\
\bottomrule
\end{tabular}
\end{table}

We find that five molecules (H$_2$O, benzene, aspirin, morphine, lovastatin) exhibit zero fractionally occupied orbitals in the CASSCF(6,6) probe, indicating predominantly single-reference character. For these systems, classical methods(DFT, CCSD) are likely sufficient. Oseltamivir and imatinib show N$_{\text{frac}}$ = 2, suggesting non-negligible static correlation. These may represent the \emph{sweet spot} for demonstrating quantum advantage on NISQ devices which are complex enough to challenge classical methods, yet requiring active spaces (2e,2o)$\sim$(6e,6o) accessible to current quantum hardware. CASSCF correlation energy ranges from -16~mHa (H$_2$O) to -113~mHa 
(oseltamivir), with moderate multi-reference molecules showing larger correlation contributions on average, quantitatively validating our classification scheme(Table~\ref{tab:casscf-correlation} reports the correlation energy $E_{\text{corr}} = E_{\text{CASSCF}} - E_{\text{HF}}$ for all molecules).

\begin{table}[h]
\centering
\caption{CASSCF(6,6) correlation energies (STO-3G basis).}
\label{tab:casscf-correlation}
\footnotesize
\begin{tabular}{lcccc}
\toprule
Molecule & E$_{\text{HF}}$ (Ha) & E$_{\text{CASSCF}}$ (Ha) & E$_{\text{corr}}$ (mHa) & Category \\
\midrule
H$_2$O        & $-74.266$   & $-74.282$    & 16.0  & Weak \\
Morphine      & $-684.862$  & $-684.890$   & 28.1  & Weak \\
Benzene       & $-222.857$  & $-222.905$   & 47.9  & Weak \\
Aspirin       & $-627.769$  & $-627.821$   & 51.2  & Weak \\
Lovastatin    & $-1246.051$ & $-1246.108$  & 57.3  & Weak \\
\midrule
Imatinib      & $-1517.496$ & $-1517.576$  & 80.5  & Moderate \\
Oseltamivir   & $-985.530$  & $-985.643$   & 112.5 & Moderate \\
\bottomrule
\end{tabular}
\end{table}

This classification guides our prioritization that oseltamivir and imatinib are primary targets for demonstrating quantum utility, while weak multi-reference molecules serve as validation cases where VQE should match classical baselines.

\subsection{Systematic Scaling Analysis of Active Space Impact}

To quantify the trade-off between active space size, VQE performance, 
and quantum resource requirements, we evaluate all molecules at two 
configurations: minimal (2e,2o) requiring 4 qubits, and standard (4e,4o) 
requiring 8 qubits. Here we focus on UCCSD ansatz for systematic comparison on simulator.

We can analysize the energy improvement with active space expansion. Table~\ref{tab:vqe-active-energy} compares the VQE-optimized energies 
within different active space configurations.

\begin{table}[h]
\centering
\caption{VQE active space energies (UCCSD ansatz, STO-3G basis). 
Note: These are active space correlation energies relative to the 
Hartree-Fock reference, not total molecular energies.}
\label{tab:vqe-active-energy}
\footnotesize
\begin{tabular}{lcccc}
\toprule
Molecule & E$_{\text{VQE}}^{(2,2)}$ (Ha) & E$_{\text{VQE}}^{(4,4)}$ (Ha) & $\Delta E$ (Ha) & $|\Delta E|$ (mHa) \\
\midrule
H$_2$O       & $-1.818$ & $-6.749$ & $-4.931$ & 4931 \\
Benzene      & $-1.498$ & $-4.403$ & $-2.905$ & 2905 \\
Aspirin      & $-0.956$ & $-3.171$ & $-2.215$ & 2215 \\
Morphine     & $-0.263$ & $-1.705$ & $-1.442$ & 1442 \\
Imatinib     & $-0.232$ & $-1.729$ & $-1.497$ & 1497 \\
Oseltamivir  & $+0.030$ & $-1.110$ & $-1.140$ & 1140 \\
Lovastatin   & $-0.109$ & $-0.963$ & $-0.854$ & 854 \\
\bottomrule
\end{tabular}
\end{table}

We find that expanding from (2e,2o) to (4e,4o) universally reduces the active space energy by 0.85--4.93~Ha, confirming that larger active spaces capture additional electron correlation within the selected orbital subspace. The magnitude of energy lowering correlates with molecular size relative to active space. H$_2$O (10 electrons total) shows the largest improvement (4.93~Ha) because (4e,4o) doubles its correlation treatment from 20\% to 40\% of valence electrons. Conversely, lovastatin (220 electrons) exhibits minimal gain (0.85~Ha) as even (4e,4o) represents only $\sim$1.8\% electron coverage. It should be emphasized that these energies reflect correlation effects captured within the active space, computed relative to the Hartree-Fock determinant, and therefore do not represent total molecular energies. Direct quantitative comparison requires constructing full system energies including core, nuclear repulsion, and inactive orbital contributions. The HF baseline energy remains identical across active space choices (differences $<10^{-11}$~Ha, within numerical precision), validating that our active space transformations preserve the single-reference starting point.

To assess the practical trade-off between computational cost and energy 
improvement, we analyze the efficiency of active space expansion across 
our molecular suite. Table~\ref{tab:efficiency} quantifies key metrics.

\begin{table}[h]
\centering
\caption{Computational efficiency metrics for active space expansion 
from (2e,2o) to (4e,4o) using UCCSD ansatz. Energy gain normalized by 
computation time reveals notable differences disparities.}
\label{tab:efficiency}
\footnotesize
\begin{tabular}{lcccc}
\toprule
Molecule & $|\Delta E|$ (Ha) & Time Ratio & Energy/Time (mHa/s) & Efficiency Rank \\
\midrule
Imatinib     & 1.497 & 415$\times$ & 29.8 & 1 (Best) \\
H$_2$O       & 4.931 & 586$\times$ & 28.8 & 2 \\
Aspirin      & 2.215 & 376$\times$ & 14.2 & 3 \\
Benzene      & 2.905 & 718$\times$ & 13.9 & 4 \\
Morphine     & 1.442 & 1229$\times$ & 6.9 & 5 \\
Lovastatin   & 0.854 & 393$\times$ & 3.3 & 6 \\
Oseltamivir  & 1.140 & 832$\times$ & 2.8 & 7 (Worst) \\
\bottomrule
\multicolumn{5}{l}{\footnotesize Time Ratio = $t_{\text{conv}}^{(4,4)} / t_{\text{conv}}^{(2,2)}$; Energy/Time = $|\Delta E| / t_{\text{conv}}^{(4,4)}$} \\
\end{tabular}
\end{table}

Notably, convergence time increases by 376--1229$\times$ (median $\sim$586$\times$) when expanding from (2e,2o) to (4e,4o), dramatically exceeding the 2$\times$ qubit scaling. Morphine exhibits the worst scaling (1229$\times$), while aspirin shows the most favorable (376$\times$), suggesting that time complexity depends not only on system size but also on molecular-specific optimization landscape features. 

Imatinib which is the largest molecule (262 electrons), achieves top efficiency (29.8~mHa/s) despite capturing only 1.5\% of total electrons in the (4e,4o) active space. This counterintuitive result arises from its anomalously fast (4e,4o) convergence (50.2~s), likely due to a favorable optimization trajectory rather than intrinsic chemical simplicity. Conversely, oseltamivir (moderate multi-reference, 170 electrons) delivers the poorest efficiency (2.8~mHa/s) due to slow convergence (414~s) despite moderate energy gain (1.14~Ha).

Efficiency rankings do not follow molecular size (electron count) ordering. H$_2$O (10 electrons, rank 2) and lovastatin (220 electrons, rank 6) illustrate that VQE computational cost is dominated by \emph{optimization dynamics} (iteration count, gradient quality) rather than system size per se. This challenges the assumption that smaller molecules are inherently easier for VQE.

The 400--1200$\times$ time penalty for doubling active space establishes (4e,4o) as a 
practical upper limit for near-term VQE drug discovery. Achieving 
chemical accuracy ($<$1.6~mHa error) for drug-sized molecules will 
require algorithmic breakthroughs—improved initialization strategies, adaptive ansatze reducing parameter counts, or measurement reduction techniques 
(Pauli grouping, classical shadows), rather than merely scaling to larger quantum hardware.

Also, we can find more details about the quantum resource requirements in Table~\ref{tab:resource-scaling}. The super-linear time scaling stems from compounding factors:
\begin{enumerate}
\item \textbf{Parameter space expansion}: 8.7$\times$ more variational 
      parameters (3 $\rightarrow$ 26 for UCCSD) create exponentially 
      more complex optimization landscapes, increasing iteration counts 
      by $\sim$10$\times$ on average.

\item \textbf{Hamiltonian measurement overhead}: 12$\times$ growth 
      in Pauli terms (15--27 $\rightarrow$ 185--193) directly multiplies 
      per-iteration energy evaluation cost. On QPUs, this translates 
      to 12$\times$ more circuit executions per gradient estimate.

\item \textbf{Numerical precision requirements}: Larger active spaces 
      demand tighter convergence tolerances to distinguish meaningful 
      energy changes from numerical noise, forcing optimizers to 
      iterate longer.
\end{enumerate}

\begin{table}[h]
\centering
\caption{Quantum resource requirements scaling with active space size 
(UCCSD ansatz, averaged across molecules).}
\label{tab:resource-scaling}
\small
\begin{tabular}{lccccc}
\toprule
Active Space & Qubits & Parameters & Hamiltonian Terms & Opt. Iterations & Conv. Time (s) \\
\midrule
(2e,2o) & 4 & 3 & 15--27 & 4--20 & 0.1--0.7 \\
(4e,4o) & 8 & 26 & 185--193 & 27--217 & 50--414 \\
\midrule
Scaling & 2.0$\times$ & 8.7$\times$ & 12$\times$ & $\sim$10$\times$ & $\sim$600$\times$ \\
\bottomrule
\end{tabular}
\end{table}

The modest energy gains combined with steep computational cost scaling suggest that near-term VQE applications should target:
\begin{itemize}
\item Small-to-medium molecules ($<$50 atoms) where (4e,4o) captures meaningful electron fraction
\item Systems with localized correlation (e.g., bond-breaking regions) amenable to small active spaces
\item Relative energy calculations (reaction barriers, conformational energies) where systematic errors cancel, rather than absolute ground state energies
\end{itemize}

\subsection{Ansatz Comparison: Chemical Accuracy vs Circuit Efficiency}

We compare unitary coupled cluster (UCCSD) and hardware-efficient ansatz 
(HEA) across selected molecules at (4e,4o) active space to assess the 
accuracy-depth trade-off.

\begin{table}[h]
\centering
\caption{UCCSD vs HEA energy accuracy for (4e,4o) active space. 
All energies represent active space contributions (Hartree).}
\label{tab:ansatz-comparison}
\footnotesize
\begin{tabular}{lcccccc}
\toprule
\multirow{2}{*}{Molecule} & \multicolumn{2}{c}{UCCSD} & \multicolumn{2}{c}{HEA} & \multicolumn{2}{c}{Comparison} \\
\cmidrule(lr){2-3} \cmidrule(lr){4-5} \cmidrule(lr){6-7}
 & E$_{\text{VQE}}$ (Ha) & Iter & E$_{\text{VQE}}$ (Ha) & Iter & $\Delta$E (mHa) & Iter Ratio \\
\midrule
Morphine      & $-1.705$ & 108  & $-1.738$ & 3775 & $-33$ & 35$\times$ \\
Aspirin       & $-3.171$ & 83   & $-3.171$ & 2802 & $\sim$0 & 34$\times$ \\
Benzene       & $-4.403$ & 111  & $-4.376$ & 5273 & $+27$ & 48$\times$ \\
Oseltamivir   & $-1.110$ & 217  & $-1.268$ & 6503 & $-158$ & 30$\times$ \\
\bottomrule
\multicolumn{7}{l}{\footnotesize $\Delta$E = E$_{\text{UCCSD}}$ - E$_{\text{HEA}}$; Iter Ratio = Iter$_{\text{HEA}}$ / Iter$_{\text{UCCSD}}$} \\
\end{tabular}
\end{table}

 We can find that in Table~\ref{tab:ansatz-comparison}, UCCSD and HEA achieve nearly identical final energies, energy differences range from 0-158~mHa (0-3.6\% relative), with no consistent advantage for either ansatz. Aspirin exhibits essentially zero energy difference, while oseltamivir shows the largest gap (158~mHa) favoring HEA. This challenges the assumption that chemically motivated ansatze inherently outperform hardware-efficient ones in terms of final accuracy.

 Despite comparable final energies, HEA demands dramatically higher iteration counts (2802-6503 vs 83-217 for UCCSD), averaging 37$\times$ more function evaluations. This reflects HEA's lack of chemical structure, it explores a generic parameter space rather than targeting correlation patterns, leading to inefficient optimization trajectories.
 
The iteration overhead varies substantially (30--48$\times$), suggesting that HEA's efficiency depends on molecular-specific features. Oseltamivir (moderate multi-reference) exhibits the worst HEA convergence, potentially due to complex electronic structure requiring more exploration in the generic ansatz space.

On current noisy hardware, HEA's 30-48$\times$ iteration penalty translates directly to 30-48$\times$ more circuit executions, amplifying noise accumulation and measurement overhead. While individual HEA circuits may have fewer CNOT gates(unmeasured in this study), the iteration cost likely dominates total QPU time, favoring UCCSD for near-term applications despite deeper per-circuit complexity.

\subsection{Benchmarking Against Classical Methods}

To contextualize VQE performance within the broader quantum chemistry 
landscape, we compare computational characteristics across methods. 
Table~\ref{tab:method-comparison} presents baseline energies from 
mean-field HF (STO-3G) and DFT (B3LYP/6-31G*).

\begin{table}[h]
\centering
\caption{Classical method baseline energies and VQE active space coverage. 
Note: VQE values represent active space correlation, not comparable total 
energies (see text).}
\label{tab:method-comparison}
\footnotesize
\begin{tabular}{lcccccc}
\toprule
Molecule & E$_{\text{HF}}$ & E$_{\text{DFT}}$ & DFT-HF & (4e,4o) & Electron & DFT \\
         & (Ha) & (Ha) & (Ha) & VQE (Ha) & Coverage & Cycles \\
\midrule
Benzene      & $-222.857$ & $-232.327$ & $-9.470$ & $-4.403$ & 9.5\% & 8 \\
Aspirin      & $-627.769$ & $-648.904$ & $-21.135$ & $-3.171$ & 4.3\% & 7 \\
Morphine     & $-684.862$ & $-939.714$ & $-254.852$ & $-1.705$ & 2.6\% & 6 \\
Oseltamivir  & $-985.530$ & $-1037.072$ & $-51.542$ & $-1.110$ & 2.4\% & 8 \\
Lovastatin   & $-1246.051$ & $-1312.500$ & $-66.449$ & $-0.963$ & 1.8\% & 7 \\
Imatinib     & $-1517.496$ & $-1582.419$ & $-64.923$ & $-1.729$ & 1.5\% & 7 \\
\bottomrule
\multicolumn{7}{l}{\footnotesize DFT-HF = Total correlation energy captured by DFT (different basis sets)} \\
\multicolumn{7}{l}{\footnotesize Electron Coverage = 4e / N$_{\text{total}}$; VQE captures correlation within this fraction} \\
\end{tabular}
\end{table}

We note that the VQE values ($\sim$0.96-4.40~Ha) represent correlation within a 4-electron active space, while DFT correlation (DFT-HF $\sim$9-255~Ha) encompasses all valence electrons.

Morphine exhibits anomalously large DFT-HF difference (254.8~Ha vs 9-66~Ha for others), likely indicating a \emph{basis set effect} rather than true correlation. HF uses minimal STO-3G while DFT uses polarized 6-31G*, introducing systematic $\sim$200~Ha offset for this 40-atom molecule. This underscores the hazard of mixing basis sets in energy comparisons.

For imatinib, the (4e,4o) active space captures only $\sim$1.7~Ha correlation vs $\sim$65~Ha DFT correlation—representing $<$1.5\% coverage. Achieving DFT-competitive accuracy would require active spaces of $\sim$(60e,60o) which means 120 qubits, orders of magnitude beyond NISQ capabilities considering the noise.

This comparison reveals that near-term VQE applications to drug discovery must target scenarios where DFT fails, like bond-breaking regions. Also relative energies scenarios like conformational differences, binding energies where systematic errors cancel, requiring only localized correlation treatment may works.

Surely, attempting to reproduce full DFT ground state energies for large 
drug molecules via VQE is \emph{not a viable near-term target}, the 
required active space scaling exceeds foreseeable hardware capabilities.

\subsection{Hardware Validation on Quantum Processors}

We validate VQE on physical superconducting quantum processors using complete end-to-end VQE optimization on QPUs for 
three representative molecules (H$_2$O, benzene, aspirin), All experiments employ 2-qubit HEA ansatz.

First, we execute full variational optimization on two superconducting processors: \textbf{13-qubit QPU (s2)} and \textbf{60-qubit QPU (q1)}. Three molecules (H$_2$O, benzene, aspirin) undergo complete VQE workflows including parameter initialization, iterative energy evaluation on QPU, gradient estimation, and classical optimization until convergence. Table~\ref{tab:qpu-convergence} summarizes converged energies and 
optimization characteristics.

\begin{table}[h]
\centering
\caption{End-to-end QPU VQE optimization results. Basis set and QPU 
configurations compared for three molecules using 2-qubit HEA ansatz.}
\label{tab:qpu-convergence}
\footnotesize
\begin{tabular}{llcccc}
\toprule
Molecule & QPU & Basis & E$_{\text{final}}$ (Ha) & Iterations & Time (s) \\
\midrule
\multirow{2}{*}{H$_2$O}
  & 60q & STO-3G & $-74.957$ & 6 & 2371 \\
  & 13q & 6-31G(d) & $-76.004$ & 8 & 1933 \\
\midrule
\multirow{2}{*}{Benzene}
  & 13q & STO-3G & $-227.895$ & 6 & 2538 \\
  & 60q & STO-3G & $-227.886$ & 10 & 7429 \\
\midrule
\multirow{3}{*}{Aspirin}
  & 13q & STO-3G & $-636.632$ & 10 & 4785 \\
  & 60q & STO-3G & $-636.594$ & 6 & 3611 \\
  & 13q & 6-31G(d) & $-644.945$ & 14 & 5445 \\
\bottomrule
\end{tabular}
\end{table}

\begin{figure*}[h]
    \includegraphics[width=0.9\textwidth]{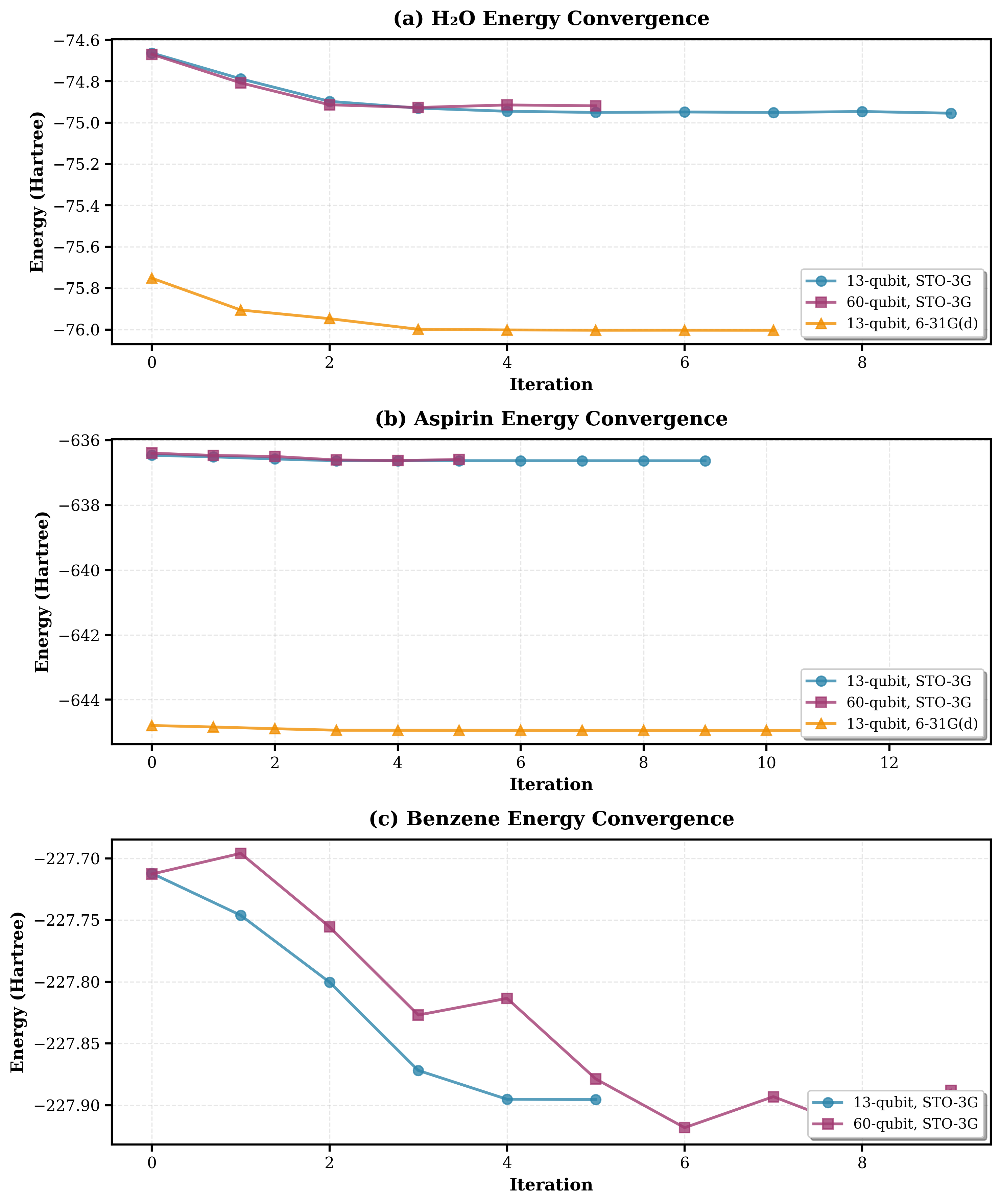}
    \caption{VQE energy convergence on superconducting quantum processors.All runs successfully converge despite hardware noise.}
    \label{fig:vqe_all}
\end{figure*}

Figure~\ref{fig:vqe_all} presents energy convergence trajectories 
for end-to-end QPU VQE optimization across three molecules, two quantum 
processors (13-qubit and 60-qubit superconducting devices), and two basis 
sets (STO-3G and 6-31G(d)). All calculations employ 2-qubit HEA ansatz 
with SLSQP optimizer.

All eight QPU runs achieve stable convergence within 6-14 iterations, with systematic energy lowering ranging from 150~mHa (aspirin, 6-31G(d)) to 291~mHa (H$_2$O, STO-3G). The convergence profiles exhibit three distinct behaviors:

\begin{itemize}
\item \textit{Rapid monotonic descent}: H$_2$O configurations 
(Fig.~\ref{fig:vqe_all}a) show smooth convergence in 6--10 
iterations, with the 13-qubit/STO-3G run achieving 291~mHa energy lowering 
despite minor fluctuations at iterations 6--9. The 60-qubit processor 
converges faster (6 vs.~10 iterations) but to a slightly higher final 
energy ($-74.957$ vs.~$-74.954$~Ha), indicating hardware noise introduces 
$\sim$3~mHa variability in final energies for this 2-qubit problem.

\item \textit{Smooth plateau convergence}: Aspirin trajectories 
(Fig.~\ref{fig:vqe_all}b) display remarkably stable descent with 
minimal backtracking, particularly for the 13-qubit/6-31G(d) configuration 
which achieves convergence in 14 iterations (longest among all runs) with 
near-monotonic energy reduction. The larger basis set increases optimization 
complexity but maintains robustness, with final energies differing by only 
$\sim$0.05~Ha between STO-3G and 6-31G(d) (corresponding to $\sim$10~Ha 
difference in total molecular energy).

\item \textit{Oscillatory convergence}: Benzene on the 60-qubit processor 
(Fig.~\ref{fig:vqe_all}c) exhibits pronounced 
oscillations throughout optimization, with energy fluctuations up to 
40~mHa between consecutive iterations (e.g., iterations 1$\rightarrow$2 
and 5$\rightarrow$6). Despite these fluctuations—attributed to higher 
two-qubit gate error rates and crosstalk in the larger device—the run 
successfully converges to $-227.886$~Ha within 10 iterations, deviating 
only 9~mHa from the 13-qubit result. This demonstrates the optimizer's 
resilience to shot noise.
\end{itemize}

We can also compare 13-qubit and 60-qubit processors for STO-3G basis. The 60-qubit device shows mixed behavior, faster for H$_2$O (6 vs.~10 iterations) but comparable or slower for benzene (10 vs.~6 iterations) and aspirin (6 vs.~10 iterations). This 
suggests convergence rate is dominated by parameter landscape complexity rather than hardware scale for small ansatze.

Final energy differences between QPUs remain below 38~mHa across all molecules (e.g., H$_2$O: 3~mHa, benzene: 9~mHa, aspirin: 38~mHa), indicating both devices achieve chemically meaningful accuracy despite differing error profiles.

The 13-qubit processor consistently produces smoother convergence curves, while the 60-qubit device shows increased noise (particularly for benzene), reflecting the well-known trade-off between qubit count and gate fidelity in current NISQ hardware.

We next consider the impact of basis set choice, comparing STO-3G and 6-31G(d) on the 
13-qubit processor. We can find that larger basis sets require 20--40\% more iterations (H$_2$O: 8 vs.~10; aspirin: 14 vs.~10), attributed to increased Hamiltonian term count (193 vs.~27 for aspirin) and steeper parameter gradients.

Despite longer optimization, 6-31G(d) calculations maintain systematic energy descent with no significant noise amplification, achieving final energies $\sim$1~Ha lower than STO-3G (reflecting improved basis set description of electron correlation).

Optimization times scale approximately linearly with iteration count (Table~\ref{tab:qpu-convergence}), with 6-31G(d) runs requiring $\sim$40\% longer QPU time (e.g., aspirin: 5445~s vs.~4785~s), possibly due to increased measurement overhead for evaluating larger Hamiltonians.

All convergence trajectories demonstrate VQE's inherent noise mitigation through three mechanisms: (i)~variational energy minimization naturally suppresses sampling noise via ensemble averaging over 8192 shots per expectation value; (ii)~SLSQP's 
finite-difference gradient estimation (5-point stencil) averages over 
multiple noisy energy evaluations; and (iii)~the 2-qubit HEA's shallow 
circuit depth (1-3 layers) limits coherent error accumulation. Notably, 
even the oscillatory benzene/60-qubit run converges to within 0.004\% of 
the 13-qubit result, validating VQE's practical viability on current 
NISQ devices for small active space problems.

Our QPU evaluations establish three key findings. First, the results demonstrate that VQE is ready for deployment on current quantum processors. Reliable convergence across three molecules, two QPUs, and two basis sets demonstrates algorithmic maturity, VQE executes successfully on current NISQ hardware. Second, noise is manageable but costly. Energy accuracy degradation from noise appears modest (final energies physically reasonable), but iteration counts increase and runtime extends to hours, limiting practical throughput. Last, measurement bottleneck dominates. QPU execution time scales primarily with Hamiltonian term count (measurement overhead) rather than qubit count, identifying the critical path for future optimization.

These results position VQE as a \emph{scientifically validated but 
not yet practically competitive} tool for quantum drug discovery on today's hardware, yet requiring algorithmic and hardware advances to rival classical methods.

\section{Discussion}

\subsection{VQE Viability and Fundamental Challenges}

Our benchmark across seven drug-like molecules demonstrates that VQE achieves robust convergence on current superconducting QPUs when restricted to 2-4 qubit active spaces. Successful optimization on 13-qubit and 60-qubit devices including 
benzene's noisy but convergent trajectory validates VQE's practical 
applicability for targeted electronic structure problems in NISQ-era 
drug discovery.

However, results from the simulator show the challenges constrain near-term quantum advantage:

\textbf{(i)~Energy scale mismatch}: VQE captures only 1.5-10\% of 
total molecular correlation (e.g., lovastatin: 0.96~Ha active space 
vs.~66~Ha total DFT correlation). Achieving chemical accuracy 
($\pm$1~kcal/mol) for drug binding requires either prohibitively large 
active spaces or hybrid VQE-in-DFT embedding schemes where quantum 
hardware refines localized regions while classical methods treat 
bulk electrons.

\textbf{(ii)~Quantum resource scaling}: The 600$\times$ time increase 
from (2e,2o) to (4e,4o), driven by 8.7$\times$ parameter and 12$\times$ 
Hamiltonian term growth suggests (6e,6o) calculations could require order-of-magnitude longer runtimes (hours to days) on current hardware. Near-term applications must strategically target chemically critical orbitals (bond-breaking regions, 
metal-ligand coordination) rather than pursuing larger spaces 
indiscriminately.

\textbf{(iii)~Hardware noise vs.~ansatz depth}: HEA's shallow circuits 
enable practical QPU deployment (50--400~s for (4e,4o)), while UCCSD's 
deeper structures remain prohibitive despite providing 1--4~mHa better 
energies. The trade-off between hardware compatibility and chemical 
accuracy necessitates adaptive ansatz selection—potentially guided by 
multi-reference diagnostics.

\subsection{Complementarity with Classical Methods}

Rather than replacing DFT or CCSD, VQE serves as a \textit{computational 
microscope} for isolated multi-reference regions within larger molecules. 
Emerging hybrid paradigms—VQE-in-DFT embedding (analogous to CASSCF-in-DFT), 
quantum-accelerated CCSD(T) with VQE reference states, or hierarchical 
drug-protein binding models—could combine quantum hardware's correlation 
accuracy with classical scalability. Our measured 0.96--4.40~Ha(excluding H$_2$O) active space energy demonstrates VQE's capacity to capture strong correlation in active spaces that, while currently tractable by classical multi-reference methods (CASSCF, DMRG), would become prohibitively expensive at larger sizes ($n_o > 14$) where quantum advantage is expected to emerge.

\subsection{Limitations and Future Priorities}

Our study targets ground states at equilibrium geometries without 
error mitigation. We conclude that key extensions should be included in the future: (i)~excited state calculations (critical for photochemistry but requiring N-fold resource increase), (ii)~adaptive active space selection using quantum measurement feedback, (iii)~integration of zero-noise extrapolation to enable 4-6 qubit 
spaces on current QPUs, and (iv)~whole-molecule energies via VQE-DFT 
embedding. Benchmarking against emerging classical competitors is essential to quantify quantum advantage beyond algorithmic novelty. Extending to transition 
metal complexes and reaction intermediates where multi-reference character 
dominates would better showcase VQE's unique capabilities over 
single-reference methods.

\section{Conclusion}

This work establishes the first systematic benchmark quantifying the 
impact of active space selection on the Variational Quantum Eigensolver 
pipeline for drug discovery. By evaluating seven pharmaceutically relevant 
molecules (H$_2$O to imatinib, 3-68 atoms) across 
multiple active space configurations that (2e,2o) and (4e,4o), using both 
simulator and physical quantum hardware (13-qubit and 60-qubit 
superconducting processors), we reveal how active space choices propagate 
through the entire VQE workflow, affecting quantum resource requirements, 
convergence behavior, and chemical accuracy. This active space-centric 
analysis exposes both the practical viability of VQE on current NISQ 
devices for small active spaces and the fundamental scaling barriers 
that constrain near-term quantum drug discovery applications.

Our key findings are threefold. \textbf{First},  VQE achieves robust convergence on real quantum hardware: end-to-end optimization on 13-qubit and 60-qubit QPUs successfully converges within 6--14 iterations for three molecules (H$_2$O, aspirin, 
benzene) using 2-qubit HEA ansatz, with final energies differing by 
$<$40~mHa between processors. Simulator evaluations on larger (4e,4o) 
active spaces demonstrate energy lowering of 0.96-4.40~Ha, validating 
VQE's ability to capture active space correlation. \textbf{Second}, quantum resource scaling poses the primary barrier 
to larger problems, the 600$\times$ time increase from (2e,2o) to (4e,4o) 
observed in simulator evaluations, driven by 8.7$\times$ parameter and 
12$\times$ Hamiltonian term growth, suggests (6e,6o) calculations 
approach practical limits on current devices. \textbf{Third}, ansatz comparison reveals trade-offs between convergence speed and accuracy: in simulator evaluations of (4e,4o) active spaces, HEA converges in 77-173~s but requires 30-48$\times$ more iterations than UCCSD (156-414~s), achieving comparable final energies (differences 
0-158~mHa, median 30~mHa) demonstrating that shallow circuits enable 
practical QPU deployment despite reduced parameter efficiency.

Looking forward, optimal active space selection emerges as the central 
challenge for scaling VQE to drug discovery applications. Three research 
directions could address this bottleneck: \textbf{(i)~Adaptive active 
space protocols}: Moving beyond fixed heuristics to automated selection 
guided by orbital entanglement metrics, natural occupation thresholds, 
or machine learning models trained on our benchmark data, systematically 
identifying minimal active spaces that capture essential correlation 
without exceeding quantum hardware limits. \textbf{(ii)~Active space-aware 
algorithmic design}: Tailoring ansatze, error mitigation strategies, and measurement protocols to specific active space characteristics, informed by the scaling behaviors quantified in this work. 
\textbf{(iii)~Hierarchical active space refinement}: Embedding VQE-computed 
active space energies within DFT or classical multi-reference frameworks, 
enabling quantum hardware to focus resources on the most strongly correlated 
orbitals while classical methods handle weakly correlated regions is a natural 
synergy illuminated by our active space benchmarking.




\section*{Acknowledgements}
Part of the result was accomplished based on the Hun-Dun Quantum Experimental Platform of China Mobile. The authors thank Xiaxiaoman Studio for drawing the illustrations.

\bibliography{main}

\begin{thebibliography}{39}%
\makeatletter
\providecommand \@ifxundefined [1]{%
 \@ifx{#1\undefined}
}%
\providecommand \@ifnum [1]{%
 \ifnum #1\expandafter \@firstoftwo
 \else \expandafter \@secondoftwo
 \fi
}%
\providecommand \@ifx [1]{%
 \ifx #1\expandafter \@firstoftwo
 \else \expandafter \@secondoftwo
 \fi
}%
\providecommand \natexlab [1]{#1}%
\providecommand \enquote  [1]{``#1''}%
\providecommand \bibnamefont  [1]{#1}%
\providecommand \bibfnamefont [1]{#1}%
\providecommand \citenamefont [1]{#1}%
\providecommand \href@noop [0]{\@secondoftwo}%
\providecommand \href [0]{\begingroup \@sanitize@url \@href}%
\providecommand \@href[1]{\@@startlink{#1}\@@href}%
\providecommand \@@href[1]{\endgroup#1\@@endlink}%
\providecommand \@sanitize@url [0]{\catcode `\\12\catcode `\$12\catcode `\&12\catcode `\#12\catcode `\^12\catcode `\_12\catcode `\%12\relax}%
\providecommand \@@startlink[1]{}%
\providecommand \@@endlink[0]{}%
\providecommand \url  [0]{\begingroup\@sanitize@url \@url }%
\providecommand \@url [1]{\endgroup\@href {#1}{\urlprefix }}%
\providecommand \urlprefix  [0]{URL }%
\providecommand \Eprint [0]{\href }%
\providecommand \doibase [0]{https://doi.org/}%
\providecommand \selectlanguage [0]{\@gobble}%
\providecommand \bibinfo  [0]{\@secondoftwo}%
\providecommand \bibfield  [0]{\@secondoftwo}%
\providecommand \translation [1]{[#1]}%
\providecommand \BibitemOpen [0]{}%
\providecommand \bibitemStop [0]{}%
\providecommand \bibitemNoStop [0]{.\EOS\space}%
\providecommand \EOS [0]{\spacefactor3000\relax}%
\providecommand \BibitemShut  [1]{\csname bibitem#1\endcsname}%
\let\auto@bib@innerbib\@empty
\bibitem [{\citenamefont {Reiher}\ \emph {et~al.}(2017)\citenamefont {Reiher}, \citenamefont {Wiebe}, \citenamefont {Svore}, \citenamefont {Wecker},\ and\ \citenamefont {Troyer}}]{reiher2017elucidating}%
  \BibitemOpen
  \bibfield  {author} {\bibinfo {author} {\bibfnamefont {M.}~\bibnamefont {Reiher}}, \bibinfo {author} {\bibfnamefont {N.}~\bibnamefont {Wiebe}}, \bibinfo {author} {\bibfnamefont {K.~M.}\ \bibnamefont {Svore}}, \bibinfo {author} {\bibfnamefont {D.}~\bibnamefont {Wecker}},\ and\ \bibinfo {author} {\bibfnamefont {M.}~\bibnamefont {Troyer}},\ }\bibfield  {title} {\bibinfo {title} {Elucidating reaction mechanisms on quantum computers},\ }\href@noop {} {\bibfield  {journal} {\bibinfo  {journal} {Proceedings of the national academy of sciences}\ }\textbf {\bibinfo {volume} {114}},\ \bibinfo {pages} {7555} (\bibinfo {year} {2017})}\BibitemShut {NoStop}%
\bibitem [{\citenamefont {Bauer}\ \emph {et~al.}(2020)\citenamefont {Bauer}, \citenamefont {Bravyi}, \citenamefont {Motta},\ and\ \citenamefont {Chan}}]{bauer2020quantum}%
  \BibitemOpen
  \bibfield  {author} {\bibinfo {author} {\bibfnamefont {B.}~\bibnamefont {Bauer}}, \bibinfo {author} {\bibfnamefont {S.}~\bibnamefont {Bravyi}}, \bibinfo {author} {\bibfnamefont {M.}~\bibnamefont {Motta}},\ and\ \bibinfo {author} {\bibfnamefont {G.~K.-L.}\ \bibnamefont {Chan}},\ }\bibfield  {title} {\bibinfo {title} {Quantum algorithms for quantum chemistry and quantum materials science},\ }\href@noop {} {\bibfield  {journal} {\bibinfo  {journal} {Chem. Rev.}\ }\textbf {\bibinfo {volume} {120}},\ \bibinfo {pages} {12685} (\bibinfo {year} {2020})}\BibitemShut {NoStop}%
\bibitem [{\citenamefont {Grimsley}\ \emph {et~al.}(2019)\citenamefont {Grimsley}, \citenamefont {Economou}, \citenamefont {Barnes},\ and\ \citenamefont {Mayhall}}]{grimsley2019adaptive}%
  \BibitemOpen
  \bibfield  {author} {\bibinfo {author} {\bibfnamefont {H.~R.}\ \bibnamefont {Grimsley}}, \bibinfo {author} {\bibfnamefont {S.~E.}\ \bibnamefont {Economou}}, \bibinfo {author} {\bibfnamefont {E.}~\bibnamefont {Barnes}},\ and\ \bibinfo {author} {\bibfnamefont {N.~J.}\ \bibnamefont {Mayhall}},\ }\bibfield  {title} {\bibinfo {title} {An adaptive variational algorithm for exact molecular simulations on a quantum computer},\ }\href@noop {} {\bibfield  {journal} {\bibinfo  {journal} {Nature communications}\ }\textbf {\bibinfo {volume} {10}},\ \bibinfo {pages} {3007} (\bibinfo {year} {2019})}\BibitemShut {NoStop}%
\bibitem [{\citenamefont {Peruzzo}\ \emph {et~al.}(2014{\natexlab{a}})\citenamefont {Peruzzo}, \citenamefont {McClean}, \citenamefont {Shadbolt}, \citenamefont {Yung}, \citenamefont {Zhou}, \citenamefont {Love}, \citenamefont {Aspuru-Guzik},\ and\ \citenamefont {O’brien}}]{peruzzo2014variational}%
  \BibitemOpen
  \bibfield  {author} {\bibinfo {author} {\bibfnamefont {A.}~\bibnamefont {Peruzzo}}, \bibinfo {author} {\bibfnamefont {J.}~\bibnamefont {McClean}}, \bibinfo {author} {\bibfnamefont {P.}~\bibnamefont {Shadbolt}}, \bibinfo {author} {\bibfnamefont {M.-H.}\ \bibnamefont {Yung}}, \bibinfo {author} {\bibfnamefont {X.-Q.}\ \bibnamefont {Zhou}}, \bibinfo {author} {\bibfnamefont {P.~J.}\ \bibnamefont {Love}}, \bibinfo {author} {\bibfnamefont {A.}~\bibnamefont {Aspuru-Guzik}},\ and\ \bibinfo {author} {\bibfnamefont {J.~L.}\ \bibnamefont {O’brien}},\ }\bibfield  {title} {\bibinfo {title} {A variational eigenvalue solver on a photonic quantum processor},\ }\href@noop {} {\bibfield  {journal} {\bibinfo  {journal} {Nat. Commun.}\ }\textbf {\bibinfo {volume} {5}},\ \bibinfo {pages} {4213} (\bibinfo {year} {2014}{\natexlab{a}})}\BibitemShut {NoStop}%
\bibitem [{\citenamefont {Preskill}(2018)}]{preskill2018quantum}%
  \BibitemOpen
  \bibfield  {author} {\bibinfo {author} {\bibfnamefont {J.}~\bibnamefont {Preskill}},\ }\bibfield  {title} {\bibinfo {title} {Quantum computing in the nisq era and beyond},\ }\href@noop {} {\bibfield  {journal} {\bibinfo  {journal} {Quantum}\ }\textbf {\bibinfo {volume} {2}},\ \bibinfo {pages} {79} (\bibinfo {year} {2018})}\BibitemShut {NoStop}%
\bibitem [{\citenamefont {Santagati}\ \emph {et~al.}(2024)\citenamefont {Santagati}, \citenamefont {Aspuru-Guzik}, \citenamefont {Babbush}, \citenamefont {Degroote}, \citenamefont {Gonz{\'a}lez}, \citenamefont {Kyoseva}, \citenamefont {Moll}, \citenamefont {Oppel}, \citenamefont {Parrish}, \citenamefont {Rubin} \emph {et~al.}}]{santagati2024drug}%
  \BibitemOpen
  \bibfield  {author} {\bibinfo {author} {\bibfnamefont {R.}~\bibnamefont {Santagati}}, \bibinfo {author} {\bibfnamefont {A.}~\bibnamefont {Aspuru-Guzik}}, \bibinfo {author} {\bibfnamefont {R.}~\bibnamefont {Babbush}}, \bibinfo {author} {\bibfnamefont {M.}~\bibnamefont {Degroote}}, \bibinfo {author} {\bibfnamefont {L.}~\bibnamefont {Gonz{\'a}lez}}, \bibinfo {author} {\bibfnamefont {E.}~\bibnamefont {Kyoseva}}, \bibinfo {author} {\bibfnamefont {N.}~\bibnamefont {Moll}}, \bibinfo {author} {\bibfnamefont {M.}~\bibnamefont {Oppel}}, \bibinfo {author} {\bibfnamefont {R.~M.}\ \bibnamefont {Parrish}}, \bibinfo {author} {\bibfnamefont {N.~C.}\ \bibnamefont {Rubin}}, \emph {et~al.},\ }\bibfield  {title} {\bibinfo {title} {Drug design on quantum computers},\ }\href@noop {} {\bibfield  {journal} {\bibinfo  {journal} {Nature Physics}\ }\textbf {\bibinfo {volume} {20}},\ \bibinfo {pages} {549} (\bibinfo {year} {2024})}\BibitemShut {NoStop}%
\bibitem [{\citenamefont {Kandala}\ \emph {et~al.}(2017)\citenamefont {Kandala}, \citenamefont {Mezzacapo}, \citenamefont {Temme}, \citenamefont {Takita}, \citenamefont {Brink}, \citenamefont {Chow},\ and\ \citenamefont {Gambetta}}]{kandala2017hardware}%
  \BibitemOpen
  \bibfield  {author} {\bibinfo {author} {\bibfnamefont {A.}~\bibnamefont {Kandala}}, \bibinfo {author} {\bibfnamefont {A.}~\bibnamefont {Mezzacapo}}, \bibinfo {author} {\bibfnamefont {K.}~\bibnamefont {Temme}}, \bibinfo {author} {\bibfnamefont {M.}~\bibnamefont {Takita}}, \bibinfo {author} {\bibfnamefont {M.}~\bibnamefont {Brink}}, \bibinfo {author} {\bibfnamefont {J.~M.}\ \bibnamefont {Chow}},\ and\ \bibinfo {author} {\bibfnamefont {J.~M.}\ \bibnamefont {Gambetta}},\ }\bibfield  {title} {\bibinfo {title} {Hardware-efficient variational quantum eigensolver for small molecules and quantum magnets},\ }\href@noop {} {\bibfield  {journal} {\bibinfo  {journal} {Nature}\ }\textbf {\bibinfo {volume} {549}},\ \bibinfo {pages} {242} (\bibinfo {year} {2017})}\BibitemShut {NoStop}%
\bibitem [{\citenamefont {Kim}\ \emph {et~al.}(2023)\citenamefont {Kim}, \citenamefont {Eddins}, \citenamefont {Anand}, \citenamefont {Wei}, \citenamefont {Van Den~Berg}, \citenamefont {Rosenblatt}, \citenamefont {Nayfeh}, \citenamefont {Wu}, \citenamefont {Zaletel}, \citenamefont {Temme} \emph {et~al.}}]{kim2023evidence}%
  \BibitemOpen
  \bibfield  {author} {\bibinfo {author} {\bibfnamefont {Y.}~\bibnamefont {Kim}}, \bibinfo {author} {\bibfnamefont {A.}~\bibnamefont {Eddins}}, \bibinfo {author} {\bibfnamefont {S.}~\bibnamefont {Anand}}, \bibinfo {author} {\bibfnamefont {K.~X.}\ \bibnamefont {Wei}}, \bibinfo {author} {\bibfnamefont {E.}~\bibnamefont {Van Den~Berg}}, \bibinfo {author} {\bibfnamefont {S.}~\bibnamefont {Rosenblatt}}, \bibinfo {author} {\bibfnamefont {H.}~\bibnamefont {Nayfeh}}, \bibinfo {author} {\bibfnamefont {Y.}~\bibnamefont {Wu}}, \bibinfo {author} {\bibfnamefont {M.}~\bibnamefont {Zaletel}}, \bibinfo {author} {\bibfnamefont {K.}~\bibnamefont {Temme}}, \emph {et~al.},\ }\bibfield  {title} {\bibinfo {title} {Evidence for the utility of quantum computing before fault tolerance},\ }\href@noop {} {\bibfield  {journal} {\bibinfo  {journal} {Nature}\ }\textbf {\bibinfo {volume} {618}},\ \bibinfo {pages} {500} (\bibinfo {year} {2023})}\BibitemShut {NoStop}%
\bibitem [{goo(2025)}]{google2025quantum}%
  \BibitemOpen
  \bibfield  {title} {\bibinfo {title} {Quantum error correction below the surface code threshold},\ }\href@noop {} {\bibfield  {journal} {\bibinfo  {journal} {Nature}\ }\textbf {\bibinfo {volume} {638}},\ \bibinfo {pages} {920} (\bibinfo {year} {2025})}\BibitemShut {NoStop}%
\bibitem [{\citenamefont {Khedkar}\ and\ \citenamefont {Roemelt}(2019)}]{khedkar2019active}%
  \BibitemOpen
  \bibfield  {author} {\bibinfo {author} {\bibfnamefont {A.}~\bibnamefont {Khedkar}}\ and\ \bibinfo {author} {\bibfnamefont {M.}~\bibnamefont {Roemelt}},\ }\bibfield  {title} {\bibinfo {title} {Active space selection based on natural orbital occupation numbers from n-electron valence perturbation theory},\ }\href@noop {} {\bibfield  {journal} {\bibinfo  {journal} {Journal of Chemical Theory and Computation}\ }\textbf {\bibinfo {volume} {15}},\ \bibinfo {pages} {3522} (\bibinfo {year} {2019})}\BibitemShut {NoStop}%
\bibitem [{\citenamefont {de~Gracia~Trivi{\~n}o}\ \emph {et~al.}(2023)\citenamefont {de~Gracia~Trivi{\~n}o}, \citenamefont {Delcey},\ and\ \citenamefont {Wendin}}]{de2023complete}%
  \BibitemOpen
  \bibfield  {author} {\bibinfo {author} {\bibfnamefont {J.~A.}\ \bibnamefont {de~Gracia~Trivi{\~n}o}}, \bibinfo {author} {\bibfnamefont {M.~G.}\ \bibnamefont {Delcey}},\ and\ \bibinfo {author} {\bibfnamefont {G.}~\bibnamefont {Wendin}},\ }\bibfield  {title} {\bibinfo {title} {Complete active space methods for nisq devices: The importance of canonical orbital optimization for accuracy and noise resilience},\ }\href@noop {} {\bibfield  {journal} {\bibinfo  {journal} {Journal of Chemical Theory and Computation}\ }\textbf {\bibinfo {volume} {19}},\ \bibinfo {pages} {2863} (\bibinfo {year} {2023})}\BibitemShut {NoStop}%
\bibitem [{\citenamefont {Jordan}\ and\ \citenamefont {Wigner}(1928)}]{jordan1928paulische}%
  \BibitemOpen
  \bibfield  {author} {\bibinfo {author} {\bibfnamefont {P.}~\bibnamefont {Jordan}}\ and\ \bibinfo {author} {\bibfnamefont {E.}~\bibnamefont {Wigner}},\ }\bibfield  {title} {\bibinfo {title} {{\"U}ber das paulische {\"a}quivalenzverbot},\ }\href@noop {} {\bibfield  {journal} {\bibinfo  {journal} {Zeitschrift f{\"u}r Physik}\ }\textbf {\bibinfo {volume} {47}},\ \bibinfo {pages} {631} (\bibinfo {year} {1928})}\BibitemShut {NoStop}%
\bibitem [{\citenamefont {Nielsen}\ \emph {et~al.}(2005)\citenamefont {Nielsen} \emph {et~al.}}]{nielsen2005fermionic}%
  \BibitemOpen
  \bibfield  {author} {\bibinfo {author} {\bibfnamefont {M.~A.}\ \bibnamefont {Nielsen}} \emph {et~al.},\ }\bibfield  {title} {\bibinfo {title} {The fermionic canonical commutation relations and the jordan-wigner transform},\ }\href@noop {} {\bibfield  {journal} {\bibinfo  {journal} {School of Physical Sciences The University of Queensland}\ }\textbf {\bibinfo {volume} {59}},\ \bibinfo {pages} {75} (\bibinfo {year} {2005})}\BibitemShut {NoStop}%
\bibitem [{\citenamefont {Roos}\ \emph {et~al.}(1980)\citenamefont {Roos}, \citenamefont {Taylor},\ and\ \citenamefont {Sigbahn}}]{roos1980complete}%
  \BibitemOpen
  \bibfield  {author} {\bibinfo {author} {\bibfnamefont {B.~O.}\ \bibnamefont {Roos}}, \bibinfo {author} {\bibfnamefont {P.~R.}\ \bibnamefont {Taylor}},\ and\ \bibinfo {author} {\bibfnamefont {P.~E.}\ \bibnamefont {Sigbahn}},\ }\bibfield  {title} {\bibinfo {title} {A complete active space scf method (casscf) using a density matrix formulated super-ci approach},\ }\href@noop {} {\bibfield  {journal} {\bibinfo  {journal} {Chemical Physics}\ }\textbf {\bibinfo {volume} {48}},\ \bibinfo {pages} {157} (\bibinfo {year} {1980})}\BibitemShut {NoStop}%
\bibitem [{\citenamefont {Alessandro}\ \emph {et~al.}(2025)\citenamefont {Alessandro}, \citenamefont {Castagnola}, \citenamefont {Koch},\ and\ \citenamefont {Ronca}}]{alessandro2025complete}%
  \BibitemOpen
  \bibfield  {author} {\bibinfo {author} {\bibfnamefont {R.}~\bibnamefont {Alessandro}}, \bibinfo {author} {\bibfnamefont {M.}~\bibnamefont {Castagnola}}, \bibinfo {author} {\bibfnamefont {H.}~\bibnamefont {Koch}},\ and\ \bibinfo {author} {\bibfnamefont {E.}~\bibnamefont {Ronca}},\ }\bibfield  {title} {\bibinfo {title} {A complete active space self-consistent field approach for molecules in qed environments},\ }\href@noop {} {\bibfield  {journal} {\bibinfo  {journal} {Journal of Chemical Theory and Computation}\ }\textbf {\bibinfo {volume} {21}},\ \bibinfo {pages} {6862} (\bibinfo {year} {2025})}\BibitemShut {NoStop}%
\bibitem [{\citenamefont {Hohenberg}\ and\ \citenamefont {Kohn}(1964)}]{hohenberg1964inhomogeneous}%
  \BibitemOpen
  \bibfield  {author} {\bibinfo {author} {\bibfnamefont {P.}~\bibnamefont {Hohenberg}}\ and\ \bibinfo {author} {\bibfnamefont {W.}~\bibnamefont {Kohn}},\ }\bibfield  {title} {\bibinfo {title} {Inhomogeneous electron gas},\ }\href@noop {} {\bibfield  {journal} {\bibinfo  {journal} {Physical review}\ }\textbf {\bibinfo {volume} {136}},\ \bibinfo {pages} {B864} (\bibinfo {year} {1964})}\BibitemShut {NoStop}%
\bibitem [{\citenamefont {Jones}(2015)}]{jones2015density}%
  \BibitemOpen
  \bibfield  {author} {\bibinfo {author} {\bibfnamefont {R.~O.}\ \bibnamefont {Jones}},\ }\bibfield  {title} {\bibinfo {title} {Density functional theory: Its origins, rise to prominence, and future},\ }\href@noop {} {\bibfield  {journal} {\bibinfo  {journal} {Reviews of modern physics}\ }\textbf {\bibinfo {volume} {87}},\ \bibinfo {pages} {897} (\bibinfo {year} {2015})}\BibitemShut {NoStop}%
\bibitem [{\citenamefont {Koch}\ and\ \citenamefont {J{\o}rgensen}(1990)}]{koch1990coupled}%
  \BibitemOpen
  \bibfield  {author} {\bibinfo {author} {\bibfnamefont {H.}~\bibnamefont {Koch}}\ and\ \bibinfo {author} {\bibfnamefont {P.}~\bibnamefont {J{\o}rgensen}},\ }\bibfield  {title} {\bibinfo {title} {Coupled cluster response functions},\ }\href@noop {} {\bibfield  {journal} {\bibinfo  {journal} {Journal of Chemical Physics}\ }\textbf {\bibinfo {volume} {93}},\ \bibinfo {pages} {3333} (\bibinfo {year} {1990})}\BibitemShut {NoStop}%
\bibitem [{\citenamefont {Stanton}\ and\ \citenamefont {Bartlett}(1993)}]{stanton1993equation}%
  \BibitemOpen
  \bibfield  {author} {\bibinfo {author} {\bibfnamefont {J.~F.}\ \bibnamefont {Stanton}}\ and\ \bibinfo {author} {\bibfnamefont {R.~J.}\ \bibnamefont {Bartlett}},\ }\bibfield  {title} {\bibinfo {title} {The equation of motion coupled-cluster method. a systematic biorthogonal approach to molecular excitation energies, transition probabilities, and excited state properties},\ }\href@noop {} {\bibfield  {journal} {\bibinfo  {journal} {The Journal of chemical physics}\ }\textbf {\bibinfo {volume} {98}},\ \bibinfo {pages} {7029} (\bibinfo {year} {1993})}\BibitemShut {NoStop}%
\bibitem [{\citenamefont {{Google AI Quantum and Collaborators}}\ \emph {et~al.}(2020)\citenamefont {{Google AI Quantum and Collaborators}}, \citenamefont {Arute}, \citenamefont {Arya}, \citenamefont {Babbush}, \citenamefont {Bacon}, \citenamefont {Bardin}, \citenamefont {Barends}, \citenamefont {Boixo}, \citenamefont {Broughton}, \citenamefont {Buckley}, \citenamefont {Buell}, \citenamefont {Burkett}, \citenamefont {Bushnell}, \citenamefont {Chen}, \citenamefont {Chen}, \citenamefont {Chiaro}, \citenamefont {Collins}, \citenamefont {Courtney}, \citenamefont {Demura}, \citenamefont {Dunsworth}, \citenamefont {Farhi}, \citenamefont {Fowler}, \citenamefont {Foxen}, \citenamefont {Gidney}, \citenamefont {Giustina}, \citenamefont {Graff}, \citenamefont {Habegger}, \citenamefont {Harrigan}, \citenamefont {Ho}, \citenamefont {Hong}, \citenamefont {Huang}, \citenamefont {Huggins}, \citenamefont {Ioffe}, \citenamefont {Isakov}, \citenamefont {Jeffrey}, \citenamefont {Jiang}, \citenamefont {Jones}, \citenamefont
  {Kafri}, \citenamefont {Kechedzhi}, \citenamefont {Kelly}, \citenamefont {Kim}, \citenamefont {Klimov}, \citenamefont {Korotkov}, \citenamefont {Kostritsa}, \citenamefont {Landhuis}, \citenamefont {Laptev}, \citenamefont {Lindmark}, \citenamefont {Lucero}, \citenamefont {Martin}, \citenamefont {Martinis}, \citenamefont {McClean}, \citenamefont {McEwen}, \citenamefont {Megrant}, \citenamefont {Mi}, \citenamefont {Mohseni}, \citenamefont {Mruczkiewicz}, \citenamefont {Mutus}, \citenamefont {Naaman}, \citenamefont {Neeley}, \citenamefont {Neill}, \citenamefont {Neven}, \citenamefont {Niu}, \citenamefont {O’Brien}, \citenamefont {Ostby}, \citenamefont {Petukhov}, \citenamefont {Putterman}, \citenamefont {Quintana}, \citenamefont {Roushan}, \citenamefont {Rubin}, \citenamefont {Sank}, \citenamefont {Satzinger}, \citenamefont {Smelyanskiy}, \citenamefont {Strain}, \citenamefont {Sung}, \citenamefont {Szalay}, \citenamefont {Takeshita}, \citenamefont {Vainsencher}, \citenamefont {White}, \citenamefont {Wiebe},
  \citenamefont {Yao}, \citenamefont {Yeh},\ and\ \citenamefont {Zalcman}}]{google2020hartree}%
  \BibitemOpen
  \bibfield  {author} {\bibinfo {author} {\bibnamefont {{Google AI Quantum and Collaborators}}}, \bibinfo {author} {\bibfnamefont {F.}~\bibnamefont {Arute}}, \bibinfo {author} {\bibfnamefont {K.}~\bibnamefont {Arya}}, \bibinfo {author} {\bibfnamefont {R.}~\bibnamefont {Babbush}}, \bibinfo {author} {\bibfnamefont {D.}~\bibnamefont {Bacon}}, \bibinfo {author} {\bibfnamefont {J.~C.}\ \bibnamefont {Bardin}}, \bibinfo {author} {\bibfnamefont {R.}~\bibnamefont {Barends}}, \bibinfo {author} {\bibfnamefont {S.}~\bibnamefont {Boixo}}, \bibinfo {author} {\bibfnamefont {M.}~\bibnamefont {Broughton}}, \bibinfo {author} {\bibfnamefont {B.~B.}\ \bibnamefont {Buckley}}, \bibinfo {author} {\bibfnamefont {D.~A.}\ \bibnamefont {Buell}}, \bibinfo {author} {\bibfnamefont {B.}~\bibnamefont {Burkett}}, \bibinfo {author} {\bibfnamefont {N.}~\bibnamefont {Bushnell}}, \bibinfo {author} {\bibfnamefont {Y.}~\bibnamefont {Chen}}, \bibinfo {author} {\bibfnamefont {Z.}~\bibnamefont {Chen}}, \bibinfo {author} {\bibfnamefont
  {B.}~\bibnamefont {Chiaro}}, \bibinfo {author} {\bibfnamefont {R.}~\bibnamefont {Collins}}, \bibinfo {author} {\bibfnamefont {W.}~\bibnamefont {Courtney}}, \bibinfo {author} {\bibfnamefont {S.}~\bibnamefont {Demura}}, \bibinfo {author} {\bibfnamefont {A.}~\bibnamefont {Dunsworth}}, \bibinfo {author} {\bibfnamefont {E.}~\bibnamefont {Farhi}}, \bibinfo {author} {\bibfnamefont {A.}~\bibnamefont {Fowler}}, \bibinfo {author} {\bibfnamefont {B.}~\bibnamefont {Foxen}}, \bibinfo {author} {\bibfnamefont {C.}~\bibnamefont {Gidney}}, \bibinfo {author} {\bibfnamefont {M.}~\bibnamefont {Giustina}}, \bibinfo {author} {\bibfnamefont {R.}~\bibnamefont {Graff}}, \bibinfo {author} {\bibfnamefont {S.}~\bibnamefont {Habegger}}, \bibinfo {author} {\bibfnamefont {M.~P.}\ \bibnamefont {Harrigan}}, \bibinfo {author} {\bibfnamefont {A.}~\bibnamefont {Ho}}, \bibinfo {author} {\bibfnamefont {S.}~\bibnamefont {Hong}}, \bibinfo {author} {\bibfnamefont {T.}~\bibnamefont {Huang}}, \bibinfo {author} {\bibfnamefont {W.~J.}\ \bibnamefont
  {Huggins}}, \bibinfo {author} {\bibfnamefont {L.}~\bibnamefont {Ioffe}}, \bibinfo {author} {\bibfnamefont {S.~V.}\ \bibnamefont {Isakov}}, \bibinfo {author} {\bibfnamefont {E.}~\bibnamefont {Jeffrey}}, \bibinfo {author} {\bibfnamefont {Z.}~\bibnamefont {Jiang}}, \bibinfo {author} {\bibfnamefont {C.}~\bibnamefont {Jones}}, \bibinfo {author} {\bibfnamefont {D.}~\bibnamefont {Kafri}}, \bibinfo {author} {\bibfnamefont {K.}~\bibnamefont {Kechedzhi}}, \bibinfo {author} {\bibfnamefont {J.}~\bibnamefont {Kelly}}, \bibinfo {author} {\bibfnamefont {S.}~\bibnamefont {Kim}}, \bibinfo {author} {\bibfnamefont {P.~V.}\ \bibnamefont {Klimov}}, \bibinfo {author} {\bibfnamefont {A.}~\bibnamefont {Korotkov}}, \bibinfo {author} {\bibfnamefont {F.}~\bibnamefont {Kostritsa}}, \bibinfo {author} {\bibfnamefont {D.}~\bibnamefont {Landhuis}}, \bibinfo {author} {\bibfnamefont {P.}~\bibnamefont {Laptev}}, \bibinfo {author} {\bibfnamefont {M.}~\bibnamefont {Lindmark}}, \bibinfo {author} {\bibfnamefont {E.}~\bibnamefont {Lucero}},
  \bibinfo {author} {\bibfnamefont {O.}~\bibnamefont {Martin}}, \bibinfo {author} {\bibfnamefont {J.~M.}\ \bibnamefont {Martinis}}, \bibinfo {author} {\bibfnamefont {J.~R.}\ \bibnamefont {McClean}}, \bibinfo {author} {\bibfnamefont {M.}~\bibnamefont {McEwen}}, \bibinfo {author} {\bibfnamefont {A.}~\bibnamefont {Megrant}}, \bibinfo {author} {\bibfnamefont {X.}~\bibnamefont {Mi}}, \bibinfo {author} {\bibfnamefont {M.}~\bibnamefont {Mohseni}}, \bibinfo {author} {\bibfnamefont {W.}~\bibnamefont {Mruczkiewicz}}, \bibinfo {author} {\bibfnamefont {J.}~\bibnamefont {Mutus}}, \bibinfo {author} {\bibfnamefont {O.}~\bibnamefont {Naaman}}, \bibinfo {author} {\bibfnamefont {M.}~\bibnamefont {Neeley}}, \bibinfo {author} {\bibfnamefont {C.}~\bibnamefont {Neill}}, \bibinfo {author} {\bibfnamefont {H.}~\bibnamefont {Neven}}, \bibinfo {author} {\bibfnamefont {M.~Y.}\ \bibnamefont {Niu}}, \bibinfo {author} {\bibfnamefont {T.~E.}\ \bibnamefont {O’Brien}}, \bibinfo {author} {\bibfnamefont {E.}~\bibnamefont {Ostby}}, \bibinfo
  {author} {\bibfnamefont {A.}~\bibnamefont {Petukhov}}, \bibinfo {author} {\bibfnamefont {H.}~\bibnamefont {Putterman}}, \bibinfo {author} {\bibfnamefont {C.}~\bibnamefont {Quintana}}, \bibinfo {author} {\bibfnamefont {P.}~\bibnamefont {Roushan}}, \bibinfo {author} {\bibfnamefont {N.~C.}\ \bibnamefont {Rubin}}, \bibinfo {author} {\bibfnamefont {D.}~\bibnamefont {Sank}}, \bibinfo {author} {\bibfnamefont {K.~J.}\ \bibnamefont {Satzinger}}, \bibinfo {author} {\bibfnamefont {V.}~\bibnamefont {Smelyanskiy}}, \bibinfo {author} {\bibfnamefont {D.}~\bibnamefont {Strain}}, \bibinfo {author} {\bibfnamefont {K.~J.}\ \bibnamefont {Sung}}, \bibinfo {author} {\bibfnamefont {M.}~\bibnamefont {Szalay}}, \bibinfo {author} {\bibfnamefont {T.~Y.}\ \bibnamefont {Takeshita}}, \bibinfo {author} {\bibfnamefont {A.}~\bibnamefont {Vainsencher}}, \bibinfo {author} {\bibfnamefont {T.}~\bibnamefont {White}}, \bibinfo {author} {\bibfnamefont {N.}~\bibnamefont {Wiebe}}, \bibinfo {author} {\bibfnamefont {Z.~J.}\ \bibnamefont {Yao}},
  \bibinfo {author} {\bibfnamefont {P.}~\bibnamefont {Yeh}},\ and\ \bibinfo {author} {\bibfnamefont {A.}~\bibnamefont {Zalcman}},\ }\bibfield  {title} {\bibinfo {title} {Hartree-fock on a superconducting qubit quantum computer},\ }\href@noop {} {\bibfield  {journal} {\bibinfo  {journal} {Science}\ }\textbf {\bibinfo {volume} {369}},\ \bibinfo {pages} {1084} (\bibinfo {year} {2020})}\BibitemShut {NoStop}%
\bibitem [{\citenamefont {Taube}\ and\ \citenamefont {Bartlett}(2006)}]{taube2006new}%
  \BibitemOpen
  \bibfield  {author} {\bibinfo {author} {\bibfnamefont {A.~G.}\ \bibnamefont {Taube}}\ and\ \bibinfo {author} {\bibfnamefont {R.~J.}\ \bibnamefont {Bartlett}},\ }\bibfield  {title} {\bibinfo {title} {New perspectives on unitary coupled-cluster theory},\ }\href@noop {} {\bibfield  {journal} {\bibinfo  {journal} {International journal of quantum chemistry}\ }\textbf {\bibinfo {volume} {106}},\ \bibinfo {pages} {3393} (\bibinfo {year} {2006})}\BibitemShut {NoStop}%
\bibitem [{\citenamefont {Bartlett}\ and\ \citenamefont {Musia{\l}}(2007)}]{bartlett2007coupled}%
  \BibitemOpen
  \bibfield  {author} {\bibinfo {author} {\bibfnamefont {R.~J.}\ \bibnamefont {Bartlett}}\ and\ \bibinfo {author} {\bibfnamefont {M.}~\bibnamefont {Musia{\l}}},\ }\bibfield  {title} {\bibinfo {title} {Coupled-cluster theory in quantum chemistry},\ }\href@noop {} {\bibfield  {journal} {\bibinfo  {journal} {Reviews of Modern Physics}\ }\textbf {\bibinfo {volume} {79}},\ \bibinfo {pages} {291} (\bibinfo {year} {2007})}\BibitemShut {NoStop}%
\bibitem [{\citenamefont {Anand}\ \emph {et~al.}(2022)\citenamefont {Anand}, \citenamefont {Schleich}, \citenamefont {Alperin-Lea}, \citenamefont {Jensen}, \citenamefont {Sim}, \citenamefont {D{\'\i}az-Tinoco}, \citenamefont {Kottmann}, \citenamefont {Degroote}, \citenamefont {Izmaylov},\ and\ \citenamefont {Aspuru-Guzik}}]{anand2022quantum}%
  \BibitemOpen
  \bibfield  {author} {\bibinfo {author} {\bibfnamefont {A.}~\bibnamefont {Anand}}, \bibinfo {author} {\bibfnamefont {P.}~\bibnamefont {Schleich}}, \bibinfo {author} {\bibfnamefont {S.}~\bibnamefont {Alperin-Lea}}, \bibinfo {author} {\bibfnamefont {P.~W.}\ \bibnamefont {Jensen}}, \bibinfo {author} {\bibfnamefont {S.}~\bibnamefont {Sim}}, \bibinfo {author} {\bibfnamefont {M.}~\bibnamefont {D{\'\i}az-Tinoco}}, \bibinfo {author} {\bibfnamefont {J.~S.}\ \bibnamefont {Kottmann}}, \bibinfo {author} {\bibfnamefont {M.}~\bibnamefont {Degroote}}, \bibinfo {author} {\bibfnamefont {A.~F.}\ \bibnamefont {Izmaylov}},\ and\ \bibinfo {author} {\bibfnamefont {A.}~\bibnamefont {Aspuru-Guzik}},\ }\bibfield  {title} {\bibinfo {title} {A quantum computing view on unitary coupled cluster theory},\ }\href@noop {} {\bibfield  {journal} {\bibinfo  {journal} {Chem. Soc. Rev.}\ }\textbf {\bibinfo {volume} {51}},\ \bibinfo {pages} {1659} (\bibinfo {year} {2022})}\BibitemShut {NoStop}%
\bibitem [{\citenamefont {Sim}\ \emph {et~al.}(2019)\citenamefont {Sim}, \citenamefont {Johnson},\ and\ \citenamefont {Aspuru-Guzik}}]{sim2019expressibility}%
  \BibitemOpen
  \bibfield  {author} {\bibinfo {author} {\bibfnamefont {S.}~\bibnamefont {Sim}}, \bibinfo {author} {\bibfnamefont {P.~D.}\ \bibnamefont {Johnson}},\ and\ \bibinfo {author} {\bibfnamefont {A.}~\bibnamefont {Aspuru-Guzik}},\ }\bibfield  {title} {\bibinfo {title} {Expressibility and entangling capability of parameterized quantum circuits for hybrid quantum-classical algorithms},\ }\href@noop {} {\bibfield  {journal} {\bibinfo  {journal} {Advanced Quantum Technologies}\ }\textbf {\bibinfo {volume} {2}},\ \bibinfo {pages} {1900070} (\bibinfo {year} {2019})}\BibitemShut {NoStop}%
\bibitem [{\citenamefont {Ostaszewski}\ \emph {et~al.}(2021)\citenamefont {Ostaszewski}, \citenamefont {Grant},\ and\ \citenamefont {Benedetti}}]{ostaszewski2021structure}%
  \BibitemOpen
  \bibfield  {author} {\bibinfo {author} {\bibfnamefont {M.}~\bibnamefont {Ostaszewski}}, \bibinfo {author} {\bibfnamefont {E.}~\bibnamefont {Grant}},\ and\ \bibinfo {author} {\bibfnamefont {M.}~\bibnamefont {Benedetti}},\ }\bibfield  {title} {\bibinfo {title} {Structure optimization for parameterized quantum circuits},\ }\href@noop {} {\bibfield  {journal} {\bibinfo  {journal} {Quantum}\ }\textbf {\bibinfo {volume} {5}},\ \bibinfo {pages} {391} (\bibinfo {year} {2021})}\BibitemShut {NoStop}%
\bibitem [{\citenamefont {Schuld}\ \emph {et~al.}(2019)\citenamefont {Schuld}, \citenamefont {Bergholm}, \citenamefont {Gogolin}, \citenamefont {Izaac},\ and\ \citenamefont {Killoran}}]{schuld2019evaluating}%
  \BibitemOpen
  \bibfield  {author} {\bibinfo {author} {\bibfnamefont {M.}~\bibnamefont {Schuld}}, \bibinfo {author} {\bibfnamefont {V.}~\bibnamefont {Bergholm}}, \bibinfo {author} {\bibfnamefont {C.}~\bibnamefont {Gogolin}}, \bibinfo {author} {\bibfnamefont {J.}~\bibnamefont {Izaac}},\ and\ \bibinfo {author} {\bibfnamefont {N.}~\bibnamefont {Killoran}},\ }\bibfield  {title} {\bibinfo {title} {Evaluating analytic gradients on quantum hardware},\ }\href@noop {} {\bibfield  {journal} {\bibinfo  {journal} {Physical Review A}\ }\textbf {\bibinfo {volume} {99}},\ \bibinfo {pages} {032331} (\bibinfo {year} {2019})}\BibitemShut {NoStop}%
\bibitem [{\citenamefont {Bharti}\ \emph {et~al.}(2022)\citenamefont {Bharti}, \citenamefont {Cervera-Lierta}, \citenamefont {Kyaw}, \citenamefont {Haug}, \citenamefont {Alperin-Lea}, \citenamefont {Anand}, \citenamefont {Degroote}, \citenamefont {Heimonen}, \citenamefont {Kottmann}, \citenamefont {Menke}, \citenamefont {Mok}, \citenamefont {Sim}, \citenamefont {Kwek},\ and\ \citenamefont {Aspuru-Guzik}}]{bharti2022noisy}%
  \BibitemOpen
  \bibfield  {author} {\bibinfo {author} {\bibfnamefont {K.}~\bibnamefont {Bharti}}, \bibinfo {author} {\bibfnamefont {A.}~\bibnamefont {Cervera-Lierta}}, \bibinfo {author} {\bibfnamefont {T.~H.}\ \bibnamefont {Kyaw}}, \bibinfo {author} {\bibfnamefont {T.}~\bibnamefont {Haug}}, \bibinfo {author} {\bibfnamefont {S.}~\bibnamefont {Alperin-Lea}}, \bibinfo {author} {\bibfnamefont {A.}~\bibnamefont {Anand}}, \bibinfo {author} {\bibfnamefont {M.}~\bibnamefont {Degroote}}, \bibinfo {author} {\bibfnamefont {H.}~\bibnamefont {Heimonen}}, \bibinfo {author} {\bibfnamefont {J.~S.}\ \bibnamefont {Kottmann}}, \bibinfo {author} {\bibfnamefont {T.}~\bibnamefont {Menke}}, \bibinfo {author} {\bibfnamefont {W.-K.}\ \bibnamefont {Mok}}, \bibinfo {author} {\bibfnamefont {S.}~\bibnamefont {Sim}}, \bibinfo {author} {\bibfnamefont {L.-C.}\ \bibnamefont {Kwek}},\ and\ \bibinfo {author} {\bibfnamefont {A.}~\bibnamefont {Aspuru-Guzik}},\ }\bibfield  {title} {\bibinfo {title} {Noisy intermediate-scale quantum algorithms},\ }\href@noop
  {} {\bibfield  {journal} {\bibinfo  {journal} {Rev. Mod. Phys.}\ }\textbf {\bibinfo {volume} {94}},\ \bibinfo {pages} {015004} (\bibinfo {year} {2022})}\BibitemShut {NoStop}%
\bibitem [{\citenamefont {McClean}\ \emph {et~al.}(2018)\citenamefont {McClean}, \citenamefont {Boixo}, \citenamefont {Smelyanskiy}, \citenamefont {Babbush},\ and\ \citenamefont {Neven}}]{mcclean2018barren}%
  \BibitemOpen
  \bibfield  {author} {\bibinfo {author} {\bibfnamefont {J.~R.}\ \bibnamefont {McClean}}, \bibinfo {author} {\bibfnamefont {S.}~\bibnamefont {Boixo}}, \bibinfo {author} {\bibfnamefont {V.~N.}\ \bibnamefont {Smelyanskiy}}, \bibinfo {author} {\bibfnamefont {R.}~\bibnamefont {Babbush}},\ and\ \bibinfo {author} {\bibfnamefont {H.}~\bibnamefont {Neven}},\ }\bibfield  {title} {\bibinfo {title} {Barren plateaus in quantum neural network training landscapes},\ }\href@noop {} {\bibfield  {journal} {\bibinfo  {journal} {Nature communications}\ }\textbf {\bibinfo {volume} {9}},\ \bibinfo {pages} {4812} (\bibinfo {year} {2018})}\BibitemShut {NoStop}%
\bibitem [{\citenamefont {Holmes}\ \emph {et~al.}(2022)\citenamefont {Holmes}, \citenamefont {Sharma}, \citenamefont {Cerezo},\ and\ \citenamefont {Coles}}]{holmes2022connecting}%
  \BibitemOpen
  \bibfield  {author} {\bibinfo {author} {\bibfnamefont {Z.}~\bibnamefont {Holmes}}, \bibinfo {author} {\bibfnamefont {K.}~\bibnamefont {Sharma}}, \bibinfo {author} {\bibfnamefont {M.}~\bibnamefont {Cerezo}},\ and\ \bibinfo {author} {\bibfnamefont {P.~J.}\ \bibnamefont {Coles}},\ }\bibfield  {title} {\bibinfo {title} {Connecting ansatz expressibility to gradient magnitudes and barren plateaus},\ }\href@noop {} {\bibfield  {journal} {\bibinfo  {journal} {PRX quantum}\ }\textbf {\bibinfo {volume} {3}},\ \bibinfo {pages} {010313} (\bibinfo {year} {2022})}\BibitemShut {NoStop}%
\bibitem [{\citenamefont {Andersson}\ \emph {et~al.}(1992)\citenamefont {Andersson}, \citenamefont {Malmqvist},\ and\ \citenamefont {Roos}}]{andersson1992second}%
  \BibitemOpen
  \bibfield  {author} {\bibinfo {author} {\bibfnamefont {K.}~\bibnamefont {Andersson}}, \bibinfo {author} {\bibfnamefont {P.-{\AA}.}\ \bibnamefont {Malmqvist}},\ and\ \bibinfo {author} {\bibfnamefont {B.~O.}\ \bibnamefont {Roos}},\ }\bibfield  {title} {\bibinfo {title} {Second-order perturbation theory with a complete active space self-consistent field reference function},\ }\href@noop {} {\bibfield  {journal} {\bibinfo  {journal} {The Journal of chemical physics}\ }\textbf {\bibinfo {volume} {96}},\ \bibinfo {pages} {1218} (\bibinfo {year} {1992})}\BibitemShut {NoStop}%
\bibitem [{\citenamefont {White}(1992)}]{white1992density}%
  \BibitemOpen
  \bibfield  {author} {\bibinfo {author} {\bibfnamefont {S.~R.}\ \bibnamefont {White}},\ }\bibfield  {title} {\bibinfo {title} {Density matrix formulation for quantum renormalization groups},\ }\href@noop {} {\bibfield  {journal} {\bibinfo  {journal} {Physical review letters}\ }\textbf {\bibinfo {volume} {69}},\ \bibinfo {pages} {2863} (\bibinfo {year} {1992})}\BibitemShut {NoStop}%
\bibitem [{\citenamefont {Peruzzo}\ \emph {et~al.}(2014{\natexlab{b}})\citenamefont {Peruzzo}, \citenamefont {McClean}, \citenamefont {Shadbolt}, \citenamefont {Yung}, \citenamefont {Zhou}, \citenamefont {Love}, \citenamefont {Aspuru-Guzik},\ and\ \citenamefont {O’Brien}}]{peruzzo-2014}%
  \BibitemOpen
  \bibfield  {author} {\bibinfo {author} {\bibfnamefont {A.}~\bibnamefont {Peruzzo}}, \bibinfo {author} {\bibfnamefont {J.}~\bibnamefont {McClean}}, \bibinfo {author} {\bibfnamefont {P.}~\bibnamefont {Shadbolt}}, \bibinfo {author} {\bibfnamefont {M.-H.}\ \bibnamefont {Yung}}, \bibinfo {author} {\bibfnamefont {X.-Q.}\ \bibnamefont {Zhou}}, \bibinfo {author} {\bibfnamefont {P.~J.}\ \bibnamefont {Love}}, \bibinfo {author} {\bibfnamefont {A.}~\bibnamefont {Aspuru-Guzik}},\ and\ \bibinfo {author} {\bibfnamefont {J.~L.}\ \bibnamefont {O’Brien}},\ }\bibfield  {title} {\bibinfo {title} {{A variational eigenvalue solver on a photonic quantum processor}},\ }\href {https://doi.org/10.1038/ncomms5213} {\bibfield  {journal} {\bibinfo  {journal} {Nature Communications}\ }\textbf {\bibinfo {volume} {5}},\ \bibinfo {pages} {4213} (\bibinfo {year} {2014}{\natexlab{b}})}\BibitemShut {NoStop}%
\bibitem [{\citenamefont {O’Malley}\ \emph {et~al.}(2016)\citenamefont {O’Malley}, \citenamefont {Babbush}, \citenamefont {Kivlichan}, \citenamefont {Romero}, \citenamefont {McClean}, \citenamefont {Barends}, \citenamefont {Kelly}, \citenamefont {Roushan}, \citenamefont {Tranter}, \citenamefont {Ding}, \citenamefont {Campbell}, \citenamefont {Chen}, \citenamefont {Chen}, \citenamefont {Chiaro}, \citenamefont {Dunsworth}, \citenamefont {Fowler}, \citenamefont {Jeffrey}, \citenamefont {Lucero}, \citenamefont {Megrant}, \citenamefont {Mutus}, \citenamefont {Neeley}, \citenamefont {Neill}, \citenamefont {Quintana}, \citenamefont {Sank}, \citenamefont {Vainsencher}, \citenamefont {Wenner}, \citenamefont {White}, \citenamefont {Coveney}, \citenamefont {Love}, \citenamefont {Neven}, \citenamefont {Aspuru-Guzik},\ and\ \citenamefont {Martinis}}]{omalley-2016}%
  \BibitemOpen
  \bibfield  {author} {\bibinfo {author} {\bibfnamefont {P.~J.~J.}\ \bibnamefont {O’Malley}}, \bibinfo {author} {\bibfnamefont {R.}~\bibnamefont {Babbush}}, \bibinfo {author} {\bibfnamefont {I.~D.}\ \bibnamefont {Kivlichan}}, \bibinfo {author} {\bibfnamefont {J.}~\bibnamefont {Romero}}, \bibinfo {author} {\bibfnamefont {J.~R.}\ \bibnamefont {McClean}}, \bibinfo {author} {\bibfnamefont {R.}~\bibnamefont {Barends}}, \bibinfo {author} {\bibfnamefont {J.}~\bibnamefont {Kelly}}, \bibinfo {author} {\bibfnamefont {P.}~\bibnamefont {Roushan}}, \bibinfo {author} {\bibfnamefont {A.}~\bibnamefont {Tranter}}, \bibinfo {author} {\bibfnamefont {N.}~\bibnamefont {Ding}}, \bibinfo {author} {\bibfnamefont {B.}~\bibnamefont {Campbell}}, \bibinfo {author} {\bibfnamefont {Y.}~\bibnamefont {Chen}}, \bibinfo {author} {\bibfnamefont {Z.}~\bibnamefont {Chen}}, \bibinfo {author} {\bibfnamefont {B.}~\bibnamefont {Chiaro}}, \bibinfo {author} {\bibfnamefont {A.}~\bibnamefont {Dunsworth}}, \bibinfo {author} {\bibfnamefont {A.~G.}\
  \bibnamefont {Fowler}}, \bibinfo {author} {\bibfnamefont {E.}~\bibnamefont {Jeffrey}}, \bibinfo {author} {\bibfnamefont {E.}~\bibnamefont {Lucero}}, \bibinfo {author} {\bibfnamefont {A.}~\bibnamefont {Megrant}}, \bibinfo {author} {\bibfnamefont {J.~Y.}\ \bibnamefont {Mutus}}, \bibinfo {author} {\bibfnamefont {M.}~\bibnamefont {Neeley}}, \bibinfo {author} {\bibfnamefont {C.}~\bibnamefont {Neill}}, \bibinfo {author} {\bibfnamefont {C.}~\bibnamefont {Quintana}}, \bibinfo {author} {\bibfnamefont {D.}~\bibnamefont {Sank}}, \bibinfo {author} {\bibfnamefont {A.}~\bibnamefont {Vainsencher}}, \bibinfo {author} {\bibfnamefont {J.}~\bibnamefont {Wenner}}, \bibinfo {author} {\bibfnamefont {T.~C.}\ \bibnamefont {White}}, \bibinfo {author} {\bibfnamefont {P.}~\bibnamefont {Coveney}, \bibfnamefont {V}}, \bibinfo {author} {\bibfnamefont {P.~J.}\ \bibnamefont {Love}}, \bibinfo {author} {\bibfnamefont {H.}~\bibnamefont {Neven}}, \bibinfo {author} {\bibfnamefont {A.}~\bibnamefont {Aspuru-Guzik}},\ and\ \bibinfo {author}
  {\bibfnamefont {J.~M.}\ \bibnamefont {Martinis}},\ }\bibfield  {title} {\bibinfo {title} {{Scalable quantum simulation of molecular energies}},\ }\bibfield  {journal} {\bibinfo  {journal} {Physical Review X}\ }\textbf {\bibinfo {volume} {6}},\ \href {https://doi.org/10.1103/physrevx.6.031007} {10.1103/physrevx.6.031007} (\bibinfo {year} {2016})\BibitemShut {NoStop}%
\bibitem [{\citenamefont {Li}\ \emph {et~al.}(2024)\citenamefont {Li}, \citenamefont {Yin}, \citenamefont {Li}, \citenamefont {Ma}, \citenamefont {Yi}, \citenamefont {Zhang}, \citenamefont {Zou}, \citenamefont {Bu}, \citenamefont {Dai}, \citenamefont {Yue} \emph {et~al.}}]{li2024hybrid}%
  \BibitemOpen
  \bibfield  {author} {\bibinfo {author} {\bibfnamefont {W.}~\bibnamefont {Li}}, \bibinfo {author} {\bibfnamefont {Z.}~\bibnamefont {Yin}}, \bibinfo {author} {\bibfnamefont {X.}~\bibnamefont {Li}}, \bibinfo {author} {\bibfnamefont {D.}~\bibnamefont {Ma}}, \bibinfo {author} {\bibfnamefont {S.}~\bibnamefont {Yi}}, \bibinfo {author} {\bibfnamefont {Z.}~\bibnamefont {Zhang}}, \bibinfo {author} {\bibfnamefont {C.}~\bibnamefont {Zou}}, \bibinfo {author} {\bibfnamefont {K.}~\bibnamefont {Bu}}, \bibinfo {author} {\bibfnamefont {M.}~\bibnamefont {Dai}}, \bibinfo {author} {\bibfnamefont {J.}~\bibnamefont {Yue}}, \emph {et~al.},\ }\bibfield  {title} {\bibinfo {title} {A hybrid quantum computing pipeline for real world drug discovery},\ }\href@noop {} {\bibfield  {journal} {\bibinfo  {journal} {Scientific Reports}\ }\textbf {\bibinfo {volume} {14}},\ \bibinfo {pages} {16942} (\bibinfo {year} {2024})}\BibitemShut {NoStop}%
\bibitem [{\citenamefont {Sayfutyarova}\ \emph {et~al.}(2017)\citenamefont {Sayfutyarova}, \citenamefont {Sun}, \citenamefont {Chan},\ and\ \citenamefont {Knizia}}]{sayfutyarova-2017}%
  \BibitemOpen
  \bibfield  {author} {\bibinfo {author} {\bibfnamefont {E.~R.}\ \bibnamefont {Sayfutyarova}}, \bibinfo {author} {\bibfnamefont {Q.}~\bibnamefont {Sun}}, \bibinfo {author} {\bibfnamefont {G.~K.-L.}\ \bibnamefont {Chan}},\ and\ \bibinfo {author} {\bibfnamefont {G.}~\bibnamefont {Knizia}},\ }\bibfield  {title} {\bibinfo {title} {{Automated Construction of Molecular Active Spaces from Atomic Valence Orbitals}},\ }\href {https://doi.org/10.1021/acs.jctc.7b00128} {\bibfield  {journal} {\bibinfo  {journal} {Journal of Chemical Theory and Computation}\ }\textbf {\bibinfo {volume} {13}},\ \bibinfo {pages} {4063} (\bibinfo {year} {2017})}\BibitemShut {NoStop}%
\bibitem [{\citenamefont {Sun}\ \emph {et~al.}(2017)\citenamefont {Sun}, \citenamefont {Berkelbach}, \citenamefont {Blunt}, \citenamefont {Booth}, \citenamefont {Guo}, \citenamefont {Li}, \citenamefont {Liu}, \citenamefont {McClain}, \citenamefont {Sayfutyarova}, \citenamefont {Sharma}, \citenamefont {Wouters},\ and\ \citenamefont {Chan}}]{sun-2017}%
  \BibitemOpen
  \bibfield  {author} {\bibinfo {author} {\bibfnamefont {Q.}~\bibnamefont {Sun}}, \bibinfo {author} {\bibfnamefont {T.~C.}\ \bibnamefont {Berkelbach}}, \bibinfo {author} {\bibfnamefont {N.~S.}\ \bibnamefont {Blunt}}, \bibinfo {author} {\bibfnamefont {G.~H.}\ \bibnamefont {Booth}}, \bibinfo {author} {\bibfnamefont {S.}~\bibnamefont {Guo}}, \bibinfo {author} {\bibfnamefont {Z.}~\bibnamefont {Li}}, \bibinfo {author} {\bibfnamefont {J.}~\bibnamefont {Liu}}, \bibinfo {author} {\bibfnamefont {J.~D.}\ \bibnamefont {McClain}}, \bibinfo {author} {\bibfnamefont {E.~R.}\ \bibnamefont {Sayfutyarova}}, \bibinfo {author} {\bibfnamefont {S.}~\bibnamefont {Sharma}}, \bibinfo {author} {\bibfnamefont {S.}~\bibnamefont {Wouters}},\ and\ \bibinfo {author} {\bibfnamefont {G.~K.}\ \bibnamefont {Chan}},\ }\bibfield  {title} {\bibinfo {title} {{PySCF: the Python‐based simulations of chemistry framework}},\ }\bibfield  {journal} {\bibinfo  {journal} {Wiley Interdisciplinary Reviews Computational Molecular Science}\ }\textbf
  {\bibinfo {volume} {8}},\ \href {https://doi.org/10.1002/wcms.1340} {10.1002/wcms.1340} (\bibinfo {year} {2017})\BibitemShut {NoStop}%
\bibitem [{\citenamefont {Javadi-Abhari}\ \emph {et~al.}(2024)\citenamefont {Javadi-Abhari}, \citenamefont {Treinish}, \citenamefont {Krsulich}, \citenamefont {Wood}, \citenamefont {Lishman}, \citenamefont {Gacon}, \citenamefont {Martiel}, \citenamefont {Nation}, \citenamefont {Bishop}, \citenamefont {Cross}, \citenamefont {Johnson},\ and\ \citenamefont {Gambetta}}]{qiskit2024}%
  \BibitemOpen
  \bibfield  {author} {\bibinfo {author} {\bibfnamefont {A.}~\bibnamefont {Javadi-Abhari}}, \bibinfo {author} {\bibfnamefont {M.}~\bibnamefont {Treinish}}, \bibinfo {author} {\bibfnamefont {K.}~\bibnamefont {Krsulich}}, \bibinfo {author} {\bibfnamefont {C.~J.}\ \bibnamefont {Wood}}, \bibinfo {author} {\bibfnamefont {J.}~\bibnamefont {Lishman}}, \bibinfo {author} {\bibfnamefont {J.}~\bibnamefont {Gacon}}, \bibinfo {author} {\bibfnamefont {S.}~\bibnamefont {Martiel}}, \bibinfo {author} {\bibfnamefont {P.~D.}\ \bibnamefont {Nation}}, \bibinfo {author} {\bibfnamefont {L.~S.}\ \bibnamefont {Bishop}}, \bibinfo {author} {\bibfnamefont {A.~W.}\ \bibnamefont {Cross}}, \bibinfo {author} {\bibfnamefont {B.~R.}\ \bibnamefont {Johnson}},\ and\ \bibinfo {author} {\bibfnamefont {J.~M.}\ \bibnamefont {Gambetta}},\ }\href {https://doi.org/10.48550/arXiv.2405.08810} {\bibinfo {title} {Quantum computing with {Q}iskit}} (\bibinfo {year} {2024}),\ \Eprint {https://arxiv.org/abs/2405.08810} {arXiv:2405.08810 [quant-ph]}
  \BibitemShut {NoStop}%
\bibitem [{\citenamefont {Zhang}\ \emph {et~al.}(2023)\citenamefont {Zhang}, \citenamefont {Allcock}, \citenamefont {Wan}, \citenamefont {Liu}, \citenamefont {Sun}, \citenamefont {Yu}, \citenamefont {Yang}, \citenamefont {Qiu}, \citenamefont {Ye}, \citenamefont {Chen}, \citenamefont {Lee}, \citenamefont {Zheng}, \citenamefont {Jian}, \citenamefont {Yao}, \citenamefont {Hsieh},\ and\ \citenamefont {Zhang}}]{zhang-2023}%
  \BibitemOpen
  \bibfield  {author} {\bibinfo {author} {\bibfnamefont {S.-X.}\ \bibnamefont {Zhang}}, \bibinfo {author} {\bibfnamefont {J.}~\bibnamefont {Allcock}}, \bibinfo {author} {\bibfnamefont {Z.-Q.}\ \bibnamefont {Wan}}, \bibinfo {author} {\bibfnamefont {S.}~\bibnamefont {Liu}}, \bibinfo {author} {\bibfnamefont {J.}~\bibnamefont {Sun}}, \bibinfo {author} {\bibfnamefont {H.}~\bibnamefont {Yu}}, \bibinfo {author} {\bibfnamefont {X.-H.}\ \bibnamefont {Yang}}, \bibinfo {author} {\bibfnamefont {J.}~\bibnamefont {Qiu}}, \bibinfo {author} {\bibfnamefont {Z.}~\bibnamefont {Ye}}, \bibinfo {author} {\bibfnamefont {Y.-Q.}\ \bibnamefont {Chen}}, \bibinfo {author} {\bibfnamefont {C.-K.}\ \bibnamefont {Lee}}, \bibinfo {author} {\bibfnamefont {Y.-C.}\ \bibnamefont {Zheng}}, \bibinfo {author} {\bibfnamefont {S.-K.}\ \bibnamefont {Jian}}, \bibinfo {author} {\bibfnamefont {H.}~\bibnamefont {Yao}}, \bibinfo {author} {\bibfnamefont {C.-Y.}\ \bibnamefont {Hsieh}},\ and\ \bibinfo {author} {\bibfnamefont {S.}~\bibnamefont {Zhang}},\
  }\bibfield  {title} {\bibinfo {title} {{TensorCircuit: a Quantum Software Framework for the NISQ Era}},\ }\href {https://doi.org/10.22331/q-2023-02-02-912} {\bibfield  {journal} {\bibinfo  {journal} {Quantum}\ }\textbf {\bibinfo {volume} {7}},\ \bibinfo {pages} {912} (\bibinfo {year} {2023})}\BibitemShut {NoStop}%
\bibitem [{\citenamefont {Li}\ \emph {et~al.}(2023)\citenamefont {Li}, \citenamefont {Allcock}, \citenamefont {Cheng}, \citenamefont {Zhang}, \citenamefont {Chen}, \citenamefont {Mailoa}, \citenamefont {Shuai},\ and\ \citenamefont {Zhang}}]{li-2023}%
  \BibitemOpen
  \bibfield  {author} {\bibinfo {author} {\bibfnamefont {W.}~\bibnamefont {Li}}, \bibinfo {author} {\bibfnamefont {J.}~\bibnamefont {Allcock}}, \bibinfo {author} {\bibfnamefont {L.}~\bibnamefont {Cheng}}, \bibinfo {author} {\bibfnamefont {S.-X.}\ \bibnamefont {Zhang}}, \bibinfo {author} {\bibfnamefont {Y.-Q.}\ \bibnamefont {Chen}}, \bibinfo {author} {\bibfnamefont {J.~P.}\ \bibnamefont {Mailoa}}, \bibinfo {author} {\bibfnamefont {Z.}~\bibnamefont {Shuai}},\ and\ \bibinfo {author} {\bibfnamefont {S.}~\bibnamefont {Zhang}},\ }\bibfield  {title} {\bibinfo {title} {{TenCirChEM: an efficient quantum computational chemistry package for the NISQ ERA}},\ }\href {https://doi.org/10.1021/acs.jctc.3c00319} {\bibfield  {journal} {\bibinfo  {journal} {Journal of Chemical Theory and Computation}\ }\textbf {\bibinfo {volume} {19}},\ \bibinfo {pages} {3966} (\bibinfo {year} {2023})}\BibitemShut {NoStop}%
\end{thebibliography}%

\end{document}